\documentclass[a4paper,11pt]{article}
\pdfoutput=1
\usepackage{jheppub}
\usepackage{color}
\usepackage{array}
\newcolumntype{C}[1]{>{\centering\arraybackslash}p{#1}}
\usepackage{slashed,dsfont}
\usepackage{bm,wasysym}

\usepackage{tabularx}
\usepackage{rotating}
\usepackage{lscape}

\newcommand{\lsim}{
\mathrel{\hbox{\rlap{\hbox{\lower4pt\hbox{$\sim$}}}\hbox{$<$}}}}

\newcommand{\gsim}{
\mathrel{\hbox{\rlap{\hbox{\lower4pt\hbox{$\sim$}}}\hbox{$>$}}}}

\renewcommand{\arraystretch}{2}

\def\tr{\text{tr}}

\allowdisplaybreaks[2]

\title{Renormalization Group Evolution of Flavour Invariants}

\author[a]{Thorsten Feldmann,}
\author[a]{Thomas Mannel,}
\author[b]{and Steffen Schwertfeger}

\affiliation[a]{Theoretische Elementarteilchenphysik, 
Universit\"at Siegen, 57068 Siegen, Germany}

\affiliation[b]{Physik Department T31,  Technische Universit\"at M\"unchen, 
85748 Garching, Germany}

\emailAdd{thorsten.feldmann@uni-siegen.de}
\emailAdd{mannel@physik.uni-siegen.de}
\emailAdd{schwertfeger@ph.tum.de}

\abstract{%
The fermion spectrum in the Standard Model (SM) exhibits
hierarchical structures between the eigenvalues of the 
Yukawa matrices which determine the fermion masses,
as well as certain hierarchical patterns in the mixing
matrix that describes weak transitions between different
fermion generations. A basis-independent description of 
the SM flavour structure can be given in terms of a complete
set of flavour invariants. 
In this paper, we construct a convenient set
of such invariants, and discuss the general form of the renormalization-group
equations. We also discuss the simplifications that arise
from exploiting hierarchies in Yukwawa couplings and mixings
which are present in the SM or its minimal-flavour violating
extensions. 
}


\note{Preprint SI-HEP-2014-15, QFET-2014-10}

\begin{document}

\maketitle

\renewcommand{\arraystretch}{1.2}


\section{Introduction}

In the Standard Model (SM) of particle physics,
the Yukawa couplings of quarks and charged leptons to the Higgs field
are the only sources of flavour structure. The singular
values of the Yukawa matrices, together with the vacuum
expectation value (VEV) of the Higgs field determine the 
fermion masses, and the relative orientation between the 
up- and down-quark Yukawa matrices results in the 
Cabibbo-Kobayashi-Maskawa (CKM) matrix, responsible for
charged flavour transitions in weak interactions.
In the quantum-field theoretical formulation of the SM,
the Yukawa matrices enter as coupling parameters in the 
Lagrange density. In the following, we will focus on 
the quark sector, where one has
\begin{align}
 -{\cal L}_{\rm Yukawa} &= 
   Y_U^{ij} \, \bar Q_L^i \widetilde H \, U_R^j 
 + Y_D^{ij} \, \bar Q_L^i H \, D_R^j  + \mbox{h.c.}
 \label{Yuk}
\end{align}
Here and in the following $Y_U$ and $Y_D$ denote the 
Yukawa matrices for up- and down-type quarks, $Q_L,U_R,D_R$
are the left-handed quark doublet and right-handed singlets,
respectively, and $H,\widetilde H$ is the Higgs field and
its $SU(2)$ conjugate. The indices  
$i,j=1\ldots n_g$ denote the quark generations/families ($n_g=3$ in the SM).

As all other couplings in the SM, after renormalization of ultraviolet divergencies,
the Yukawa matrices in (\ref{Yuk}) are to be interpreted as effective parameters
with the scale-dependence controlled by renormalization-group (RG) equations
\cite{Machacek:1983fi,Grzadkowski:1987tf,Arason:1991ic,Barger:1992pk}.
The structure of the RG equations and their solutions have been extensively 
studied in the past. In  \cite{Balzereit:1998id}, the resulting one-loop RG evolution 
of the CKM matrix elements (in a given parametrization) has been studied,
and approximate analytic solutions have been derived on the basis of the 
observed hierachies in quark masses and mixing angles in the SM.
Generalizations to particular new physics (NP) frameworks have also been
derived, notably for 2-Higgs-doublet models or supersymmetric extensions
of the SM, see, for instance, \cite{Das:2000uk,Crivellin:2008mq}.
Recently, the effect of possible NP contributions has 
been studied in a model-independent way, by
 considering the RG effects from dimension-six operators in 
 an effective field theory (SM-EFT) approach \cite{Jenkins:2013wua}.
Finally,
Bednyakov et al.\ \cite{Bednyakov:2014pia}
have recently computed the three-loop RG coefficients 
for the SM Yukawa matrices.
  
The RG equations are usually formulated in matrix form,
i.e.\ the scale-variation of the Yukawa matrices is 
given by a matrix polynomial of $Y_U$ and $Y_D$. Since 
the gauge sector of the SM is invariant under unitary
field redefinitions for the individual quark multiplets,
the RG equations have to transform covariantly under 
such changes of flavour basis (see below). This also 
implies a certain degree of redundancy 
in the RG equations, because from the 
18 complex matrix entries in $Y_U$ and $Y_D$ only 
10 physical parameters are observables.

In this paper, we will therefore reformulate the RG equations 
in terms of \emph{flavour invariants}, i.e.\ objects 
constructed from $Y_U$ and $Y_D$ which are independent of the 
choice of flavour basis. As has been shown in
\cite{Jenkins:2009dy} from the basic algebraic principle
of Hilbert series, one can define \emph{eleven} polynomially
independent flavour invariants for three quark generations.
These fix the six quark masses, the three mixing angles and 
the sine and cosine of the CP-violating phase in the CKM
matrix. 
As a corollary, using Cayley-Hamilton identities for matrix products
(cf.\ \cite{Colangelo:2008qp}), 
this also implies that any flavour-covariant product of Yukawa
matrices that appears on the right-hand side of the 
RG equations for $Y_U$ or $Y_D$ can be reduced to a finite basis of flavour matrices
with coefficients given as polynomials of flavour invariants.
It is then a straightforward, though tedious, task to 
derive the RG equations for the set of flavour invariants.

Although the RG equations for the flavour invariants contain the
same information as the original flavour-covariant equations,
the formulation in terms of flavour invariants, under certain
circumstances, may be considered advantageous. For instance,
the form of the RG equations is universal, not only for the SM,
but also for all extensions that obey the principle of minimal
flavour violation (MFV) in the technical sense of \cite{D'Ambrosio:2002ex}.
Furthermore, the hierachical pattern of masses and mixing directly translates
into a well-defined power counting for (suitably chosen) flavour invariants,
which can be exploited to simplify the RG equations. 
An attractive physical picture arises if one assumes these hierarchies 
to be associated to some dynamical NP mechanism that can be traced 
back to an effective potential which determines the flavour structures
at low energies. Within the MFV framework, the potential itself will have to be
formulated in terms of flavour invariants, and the minimization of the potential
should generate VEVs for the flavour invariants that reflect the 
particular pattern of (sequential) flavour-symmetry breaking in the SM
(see  \cite{Feldmann:2009dc}).
Recent studies along these lines can be found, for instance, in 
\cite{Alonso:2011yg,Nardi:2011st,Espinosa:2012uu,Fong:2013dnk}.
Finally, our approach could be extended and generalized to cases where 
there are additional flavour structures in some tensor representation of 
the SM flavour symmetry group. For example, these could show up as 
coupling constants in front of higher-dimensional operators in SM-EFT 
\cite{Buchmuller:1985jz,Grzadkowski:2010es}, or as new
spurion fields in the MFV framework \cite{Feldmann:2006jk,Isidori:2012ts}.

The remaining paper is organized as follows:
In the next section, we will first discuss a toy scenario
with only two generations (2G) of SM quarks. 
The simplifications in the 2G case (no CP violation,
closure of $SU(2)$ matrices under multiplication, 
small number of polynomially independent invariants)
allow us to introduce our approach in a very transparent
way, perform all calculations analytically and illustrate 
the RG equations for the flavour invariants in a graphical
way. To this end, we will first give convenient definitions 
for flavour invariants and basic flavour matrices. In
terms of these, the general form of the RG equations for
flavour invariants will be derived. We also present analytical and 
numerical solutions for the RG equations that can be obtained from 
exploiting SM-like flavour hierarchies in the one-loop approximation.
In Section~\ref{sec:3G} we generalize our framework to the
realistic case of three quark generations (3G). To keep the
discussion transparent, we restrict ourselves to the one-loop
approximation from the very beginning. Again, we derive the 
general form of the (one-loop) RG equations for the eleven 
flavour invariants, and discuss their approximate solutions 
in the SM.
We close the paper with a short summary and outlook
in Section~\ref{summary}.
Some technical details about the use of Cayley-Hamilton identities,
the explicit form of the 3G flavour invariants, and the general
form of the two-loop RG equations in the 3G case can be found 
in the appendices.

\section{Two Quark Generations}

As mentioned above, in this section, we restrict ourselves to two generations of
left-handed quark doublets and right-handed up- and down-quark
singlets in the SM.
The gauge-kinetic terms of the SM Lagrangian are flavour-blind,
and therefore independent unitary rotations of the 
quark multiplets define a flavour symmetry,
\begin{align}
 {\cal G}_{\rm quark} &= U(2)^3/U(1)_B \ \sim \ SU(2)_{Q_L} \times U(2)_{U_R} \times U(2)_{D_R} \,,
 \label{G2G}
\end{align}
which is only broken by the Yukawa couplings in (\ref{Yuk}).
Here we factored out a $U(1)_B$ symmetry for baryon number conservation, which is unaffected
by the SM Yukawa interactions \cite{Feldmann:2009dc}. (More precisely, we find it convenient 
to factor out a $U(1)_{Q_L}$ transformation acting on the left-handed doublets only).
In a particular flavour basis, the Yukawa matrices for up- and down-type quarks
read
\begin{align}
 Y_U = \left( \begin{array}{cc} y_u & 0 \\ 0 & y_c \end{array} \right) \,,
& \qquad
 Y_D = V_{\rm Cabbibo}  \left( \begin{array}{cc} y_d & 0 \\ 0 & y_s \end{array} \right)
=
\left( \begin{array}{cc} \cos\theta & \sin\theta \\ -\sin\theta & \cos\theta \end{array} \right)
       \left( \begin{array}{cc} y_d & 0 \\ 0 & y_s \end{array} \right) \,.
\end{align}
Under a change of flavour basis, the Yukawa matrices transform as
\begin{align}
 Y_U  &  \to V_{Q_L} \, Y_U \, V_{U_R}^\dagger \,, \qquad  Y_D   \to V_{Q_L} \, Y_D \, V_{D_R}^\dagger \,,
\end{align}
where $V_{Q_L} \in SU(2)_{Q_L}$ etc.

\subsection{Flavour Invariants}

In the 2G case, one can construct five flavour invariants that
are polynomially independent (see e.g.~\cite{Jenkins:2009dy}
and references therein for the mathematical background).
In the following, to set the stage for the 3G case to be discussed 
in Sec.~\ref{sec:3G}, we will discuss the construction
and properties of these flavour invariants step by step.

From the Yukawa matrices $Y_U$ and $Y_D$, one can construct 
flavour invariants in terms of traces or determinants of 
matrix products constructed from the 
non-negative hermitian matrices
\begin{align}
  U = Y_U Y_U^\dagger \,, \qquad D=Y_D Y_D^\dagger \,.
\end{align}
These transform as $V_{Q_L} U \, V_{Q_L}^\dagger$ and $V_{Q_L} D \, V_{Q_L}^\dagger$
under basis tranformations for the left-handed quark doublets.
A convenient choice for non-negative invariants is
\begin{align} 
&  I_1 \equiv \tr[U] = y_u^2+y_c^2 \geq 0 \,, &&  I_2 \equiv \tr[D] = y_d^2+y_s^2 \geq 0 \,, 
\cr 
& \widehat I_3 \equiv \det[U] = y_u^2 y_c^2 \geq 0 && \widehat I_4 \equiv \det[D] = y_d^2 y_s^2\geq 0 \,,
\label{positivity-2G-raw-a}
\end{align}
and 
\begin{align}
\widehat I_5 \equiv  \tr[UD] = (y_c^2 y_s^2 + y_d^2 y_u^2) \cos^2\theta 
 + (y_c^2 y_d^2 + y_s^2 y_u^2) \sin^2\theta \geq 0
  \,.
  \label{positivity-2G-raw-b}
\end{align}
Apart from discrete ambiguties related to renaming the original quark fields
in the flavour eigenbasis, they determine the four eigenvalues
for the Yukawa couplings and the Cabibbo mixing angle.
Invariants built from traces of higher powers of $U$ and $D$ are related to the above
via Cayley-Hamilton identities (see appendix~\ref{app:CH} and e.g.\ the discussion in \cite{Colangelo:2008qp}). 
Still, for the following discussion, we further 
define the polynomially \emph{dependent} invariants
\begin{align}
I_3 \equiv \frac{I_1^2}{2} - 2 \widehat I_3 = \frac12 (y_c^2- y_u^2)^2\,, \qquad 
I_4 \equiv \frac{I_2^2}{2} - 2 \widehat I_4 = \frac12 (y_s^2-y_d^2)^2 \,, 
\end{align}
and 
\begin{align}
I_5 &\equiv  \widehat I_5 - \frac{I_1 I_2}{2} = \frac12 (y_c^2-y_u^2)(y_s^2-y_d^2) \cos(2\theta) \,, \cr
\widetilde I_5 &\equiv I_3 I_4 - I_5^2 = \frac14(y_c^2- y_u^2)^2 (y_s^2-y_d^2)^2 \sin^2(2\theta) \,.
\end{align}

\paragraph{Triplet Matrices and Triplet Invariants:}

It is further convenient to divide the matrices $U$ and $D$
into singlet and triplet components with respect to the flavour group factor $SU(2)_{Q_L}$,
\begin{align}
 U &= \frac12 \, \tr[U] \, \mathds{1} + U_3 \,,
 \qquad
 D = \frac12 \, \tr[D] \, \mathds{1} + D_3 \,,
\end{align}
A third independent triplet matrix can be defined as
\begin{align}
 A_3 &= \frac{i}2 \left[ U_3, D_3 \right] \,.
\end{align}

\paragraph{Polynomial Basis:}

For generic Yukawa entries, any $2\times 2$ matrix $M$ 
that transforms as $V_{Q_L} \, M \, V_{Q_L}^\dagger$ under
$SU(2)_{Q_L}$ can be written as a finite polynomial 
of matrices from the set $\{ {\mathds 1},U_3,D_3,A_3 \}$.
For those matrices, the following multiplication
tables for symmetric and anti-symmetric products
of matrices holds.

\begin{center} 
 \vspace{1em}
\begin{tabular}{|c||c|c|c|}
\hline 
$ \{\ , \ \}$  &   $U_3$ & $D_3$ & $A_3$
                    \\
   \hline \hline 
   $U_3$  & $ I_3 \, \mathds{1} $ & $I_5 \, \mathds{1}$ & $0$
   \\
   \hline 
   $D_3$  & $I_5 \, \mathds{1}$ & $I_4 \, \mathds{1}$ & $0$  
\\ \hline 
 $A_3$ & $0$ & $0$ & $\frac12 \widetilde I_5 \, \mathds{1} $
 \\
 \hline 
 \end{tabular} \qquad 
\begin{tabular}{|c||c|c|c|}
\hline  $i \, [\ ,\ ]$  &   $U_3$ & $D_3$ & $A_3$
                    \\
   \hline \hline 
   $U_3$  & $0$ & $\phantom{-}2 \bar A_3$ & $-2\bar D_3$
   \\
   \hline 
   $D_3$  & $- 2 \bar A_3$ & $0$ & $\phantom{-}2\bar U_3$  
\\ \hline 
 $A_3$ & $\phantom{-}2 \bar D_3$ & $-2 \bar U_3$ & $0$
 \\
 \hline
 \end{tabular}
 \vspace{1em}
 \end{center}
 
\noindent
This explicitly shows that the set  $\{ {\mathds 1},U_3,D_3,A_3 \}$
closes under matrix multiplication with prefactors that are 
polynomials of the flavour invariants . 
Here we defined the  \emph{dual} matrices 
\begin{align}
 \bar U_3 \equiv \frac{i}{2} \left[D_3, A_3\right ] 
 &= \frac{I_4}{2} \, U_3 - \frac{I_5}{2} \, D_3 
 \,,\qquad 
 \bar D_3 \equiv \frac{i}{2} \left[A_3, U_3\right] 
 = \frac{I_3}{2} \, D_3 - \frac{I_5}{2} \, U_3 \,,
\end{align}
and 
\begin{align}
 \bar A_3 &\equiv \frac{i}{2} \left[U_3,D_3\right]  = A_3 \,,
\end{align}
which can be obtained from the inverse of the metric
\begin{align}
 G_{XY} = {\rm tr}[X_3Y_3] = \left( \begin{array}{ccc}
 I_3 & I_5 & 0 \\
 I_5 & I_4 & 0 \\
 0 & 0 & \frac{\widetilde I_5}{2} \end{array} \right)\qquad 
\mbox{($X,Y = U,D,A$),} 
\end{align}
as 
\begin{align}
  \left( \begin{array}{c} \bar U_3 \\ \bar D_3 \\ \bar A_3 \end{array} \right) 
  &= \frac{\widetilde I_5}{2} \, G^{-1} \, \left( \begin{array}{c} U_3\\  D_3 \\ A_3 \end{array} \right) 
  \,.
\end{align}
From this we can read off the orthogonality relations between
triplet matrices and their dual,
\begin{align}
\tr\left[X_3 \bar Y_3 \right] = \frac{\widetilde I_5}{2} \, \delta_{XY} \qquad 
\mbox{($X,Y = U,D,A$).}
\end{align}
This can be used, for instance, to decompose 
a generic $SU(2)_{Q_L}$ triplet matrix as
\begin{align}
  M_{3} &=  \sum_{X=U,D,A} \frac{2 \, \tr[\bar X_3 M_{3}]}{\widetilde I_5} \, X_3  \,.
\end{align}
Similarly, a matrix $M_U$ that transforms as a bi-doublet under 
$SU(2)_{Q_L} \times SU(2)_{U_R}$ can be decomposed as
\begin{align}
 M_U &= 
  \frac{\tr[M_U Y_U^{-1}]}{2} \, \mathds{1}  
+ \sum_{X=U,D,A} \frac{2 \, \tr[\bar X_3 M_{U}Y_U^{-1}]}{\widetilde I_5} \, X_3 
\,,
\end{align}
and analogously for $U_R \leftrightarrow D_R$. Higher tensor representations 
of the flavour symmetry group and their expansion can 
be constructed from $M_{3,U,D}$.
Notice that for \emph{generic} matrices $M_{3,U,D}$, the coefficients in these expansions
are enhanced by $(\widetilde I_5)^{-1}$ and $Y_{U,D}^{-1}$, respectively.
In contrast, the MFV hypothesis assumes these coefficients to be of order 1 or smaller
(see again \cite{Colangelo:2008qp}).

\subsection{Renormalization-Group Equations}

\subsubsection{General Form}

The Yukawa matrices are subject to renormalization-group (RG) evolution.
The generic form for the RG-running of the Yukawa matrices 
$Y_{U,D}$ can be written in manifestly flavour-symmetric form 
(see e.g.\ \cite{Grzadkowski:1987tf}). 
Using the generic decomposition into basis matrices as discussed above, we thus write 
\begin{align}
 \frac{d Y_U(\mu)}{d\ln\mu} &= 
\left( a_0(I_i,\mu) \, \mathds{1} + a_1(I_i,\mu) \, U_3 + a_2(I_i,\mu)  \, D_3+ i \, a_3(I_i,\mu) \, A_3 \right) Y_U(\mu) \,,
\cr 
\frac{d Y_D(\mu)}{d\ln\mu} &= 
\left( b_0(I_i,\mu) \, \mathds{1}  + b_1(I_i,\mu) \, D_3 + b_2(I_i,\mu) \, U_3 - i \, b_3(I_i,\mu) \, A_3 \right) Y_D(\mu) \,.
\label{2Grge:gen}
\end{align}
Each of the coefficients $a_i,b_i$ depends on flavour invariants which arise
from loop diagrams including additional Higgs-Yukawa couplings. (At one-loop accuracy,
only terms at most quadratic in the Yukawa couplings can appear within
the round brackets etc.) In the SM (or, in general, in constrained MFV models
without additional sources of CP violation),
the coefficients will be \emph{real} polynomials of the flavour invariants.\footnote{Furthermore,
if weak isospin-violating corrections are neglected, the coefficients $a_i$ and $b_i$ will be related, see e.g.\
\cite{Arason:1991ic,JuarezWysozka:2002kx}.} 
This immediately translates into RG equations for the matrices $U$ and $D$,
an from this we obtain
\begin{align}
 \frac{dI_1}{d\ln\mu} &= \tr\left[\frac{dU}{d\ln\mu}\right] = 
 2 \left(a_0 \, I_1 + a_1 \, I_3 + a_2 \, I_5 \right) \,,
 \label{dI1}
 \\
 \frac{dI_2}{d\ln\mu} &= \tr\left[\frac{dD}{d\ln\mu}\right]
 = 2 \left(b_0 \, I_2 + b_1 \, I_4 + b_2 \, I_5 \right)\,.
 \label{dI2}
\end{align}

In a similar way, one obtains the RG equations for the remaining invariants in a straightforward manner.
The RG equations for the invariants $\widehat I_3$, $\widehat I_4$ take 
a particularly simple form
\begin{align}
\frac{d \widehat I_3}{d\ln\mu} &= 4 a_0 \, \widehat I_3 \,, \qquad 
\frac{d \widehat I_4}{d\ln\mu}  = 4 b_0 \, \widehat I_4 \,.
\label{dI3hat}
\end{align}
For the invariant $I_5$, we obtain
\begin{align}
\label{dI5}
 \frac{d  I_5}{d\ln\mu} &=
 \left(2 a_0 + 2 b_0 + a_1 I_1 + b_1 I_2 \right) I_5 + a_2 I_1 I_4 + b_2 I_2 I_3 + (a_3 + b_3) \, \widetilde I_5 \,,
\end{align}
and for the invariant $\widetilde I_5$, we get
\begin{align}
\label{dI5t}
 \frac{d \widetilde I_5}{d\ln\mu} &=
\left(4a_0 +  4b_0 + 2a_1 I_1 + 2b_1 I_2 - 2(a_3+b_3) \, I_5 \right) \widetilde I_5 \,,
\end{align}

\begin{figure}[t!!pbh]
 \begin{center}
  \includegraphics[width=0.55\textwidth]{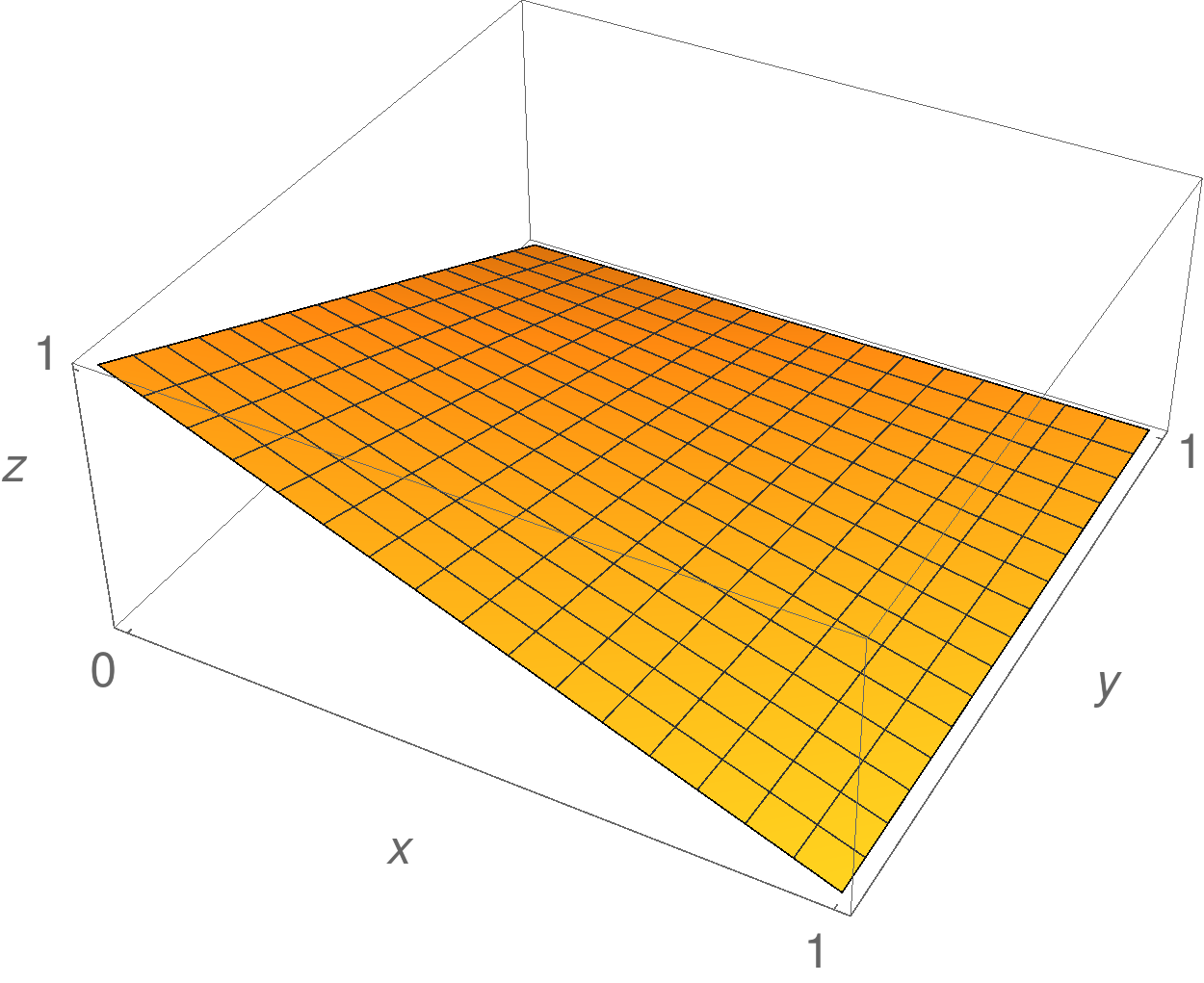}
 \end{center}
\caption{Illustration of the ``phase space'' (the region below the shaded area)
for the normalized invariants $x,y,z$ defined in the text.\label{fig:ps}}
\end{figure}

\paragraph{Discussion:} \label{subsec:disc}
From (\ref{dI3hat}) and (\ref{dI5t}) we observe that some limiting 
cases in the phase space of flavour invariants are stable under RG evolution:
\begin{itemize}
 \item \underline{The case $\widehat I_3 =0$:} \\
 In terms of physical parameters, this  corresponds to
 one vanishing eigenvalue in the up-quark sector, $y_u=0$, 
 and otherwise generic values for $y_{d,s,c}$ and $\theta$. 

 \item \underline{The case $\widehat I_4 =0$:} \\
 This corresponds to one vanishing eigenvalue in the down-quark sector, $y_d=0$, 
 and otherwise generic values for $y_{u,s,c}$ and $\theta$.
 
 \item \underline{The case $\widetilde I_5=0$;} \\
 This corresponds to $\sin2\theta=0$, i.e.\ no mixing and otherwise generic $Y_U$ and $Y_D$;
 or degenerate eigenvalues in the up-quark sector ($y_u=y_c$) or 
 in the down-quark sector ($y_d=y_s$), respectively. 
\end{itemize}
For illustration, we thus define normalized invariants (for $I_{1,2}\neq 0$), 
\begin{align}
& x =1- \bar x  \equiv \frac{4 \widehat I_3}{I_1^2} \,, \qquad 
  y= 1-\bar y  \equiv \frac{4 \widehat I_4}{I_2^2} \,, \qquad 
  z = 1-\bar z \equiv \frac{4 \widetilde I_5}{I_1^2 I_2^2}  \,,
\end{align}
which take values in the unit interval $[0,1]$,
with the additional constraints
\begin{align}
&  z  \leq \bar x \, \bar y \,.
\end{align}
This is illustrated in Fig.~\ref{fig:ps}. As a consequence of the above observations,
there will be now RG flow from the ``phase-space'' edges,
defined by $x=0$, $y=0$, or $z=0$, into the bulk. This can be understood as 
a consequence of a residual flavour symmetry. In contrast, the 
case $z=\bar x \bar y$ is not protected by symmetry.
A more detailed discussion of the residual flavour symmetries associated with this
situation will be given in \cite{Feldmann:2015unp} (see also \cite{Alonso:2013dba}).

\subsubsection{Exploiting Flavour Hierarchies}

The RG equations simplify when one exploits flavour hierarchies
in the Yukawa matrices. For instance, in a SM-like scenario, we
can consider the limit where all but one Yukawa coupling, say 
$y_c$ in the 2G toy case, are small. In this case, the 
basis of triplet matrices in 
(\ref{2Grge:gen}) can be reduced to $U_3$, and 
consequently only the coefficients $a_0,b_0,a_1,b_2$ are relevant to first approximation.\footnote{A 
complementary approach would perform the limit $y_c\gg y_{u,d,s}$ 
from the very beginning and consider invariants under the 
\emph{reduced} flavour symmetry only, see \cite{Feldmann:2008ja,Kagan:2009bn}.} 
The RG equations for the invariants in this approximation read (also
using $I_3 \simeq I_1^2/2$)
\begin{align}
  \frac{dI_1}{d\ln \mu} &\simeq \left( 2 a_0 + a_1 I_1 \right) I_1 \,,\qquad 
  \frac{d\widehat I_3}{d\ln \mu} = 4 a_0\widehat I_3   \,,
  \cr 
  \frac{dI_2}{d\ln \mu} &\simeq 2 b_0 I_2 + 2 b_2 I_5  \,, \qquad 
  \frac{d\widehat I_4}{d\ln \mu} = 4 b_0\widehat I_4   \,,
  \cr 
  \frac{d\widetilde I_5}{d\ln \mu} & \simeq \left( 4 a_0 + 4 b_0 + 2 a_1 I_1 \right) \widetilde I_5 \,.
\end{align}
Solving for the four coefficients, leaves one general relation between
the five invariants and their derivatives which can be written as
\begin{align}
\frac{d\widetilde I_5}{\widetilde I_5} &\simeq2 \, \frac{ dI_1}{I_1} + \frac{d\widehat I_4}{\widehat I_4} \qquad 
\mbox{(for $y_c \gg y_{u,d,s}$).}
\end{align}
This implies
\begin{align}
 \frac{\widetilde I_5(\mu)}{\widetilde I_5(\mu_0)} &\simeq  \frac{I_1^2(\mu) \, \widehat I_4(\mu)}{I_1^2(\mu_0) \, \widehat I_4(\mu_0)} \quad 
 \Leftrightarrow \quad \frac{y}{z} \simeq \text{const.} \,,
 \label{yz}
\end{align}
or, in terms of Yukawa eigenvalues and the Cabibbo angle,\footnote{In models with
texture zeros one typically relates the Cabibbo angle to the \emph{square root}
of $y_d/y_s$, see e.g.\ \cite{Fritzsch:1979zq}. Therefore, such relations -- 
in general -- are not scale invariant
in the limit of hierarchical Yukawa couplings.}
\begin{align}
\left( y_s/y_d-y_d/y_s \right)^2 \sin^2(2\theta)\simeq \mbox{const.}\qquad 
\mbox{(for $y_c \gg y_{u,d,s}$).} 
\end{align}
Putting in experimental values for the quark-mass ratio and the Cabibbo angle,
the constant on the r.h.s.\ ranges between 60 and 90.

\subsubsection{One-loop Solutions in the SM}

\begin{table}[t!!!pbh]
 \begin{center}
  \begin{tabular}{c|c|c|c}
   \hline\hline 
   $a_0$ & $b_0$ & $a_1=b_1$ & $a_2=b_2$  \\
   \hline 
   $\frac{3}{8\pi^2} \left(I_1+I_2 + \frac{I_1-I_2}{2n_g}  \right) - \frac{2\alpha_s}{\pi} $ &
    $\frac{3}{8\pi^2} \left(I_1+I_2 + \frac{I_2-I_1}{2n_g} \right) - \frac{2\alpha_s}{\pi} $ &
    $\frac{3}{16\pi^2}$ & $-\frac{3}{16\pi^2}$
   \\
   \hline\hline
  \end{tabular}
 \end{center}
\caption{\label{tab:aibi} Non-vanishing 
  RG  coefficients in the SM as defined in the text, following from \cite{Balzereit:1998id}.
  Here $n_g$ denotes the number of generations, and electroweak contributions have
  been neglected for simplicity. }
\end{table}

To illustrate the numerical 
effect of the RG equations, we 
consider the one-loop RG coefficients in the SM.
The system of RG equations further simplifies if we 
neglect electroweak corrections, leading to the values 
summarized in Table~\ref{tab:aibi}. 
For the starting values of the invariants in the 2G case,
we consider a toy model where we neglect
the first generation in the SM, such that the large Yukawa
couplings from the third generation lead to non-trivial 
effects on the r.h.s.\ of the RG equations.
Exploiting again the hierarchies in the SM Yukawa entries, we then find
\begin{align}
\label{I1SM}
 \frac{dI_1}{d\ln\mu} &\simeq \frac{9}{8\pi^2} \, I_1^2 - \frac{4\alpha_s}{\pi} \, I_1 \,.
\end{align}
Using the one-loop expression for the QCD $\beta$-function,
\begin{align}
\frac{d\alpha_s}{d\ln\mu} &\simeq -\frac{\beta_0}{2\pi} \, \alpha_s^2 \,,
\end{align}
one obtains the explicit solution
\begin{align}
 I_1(\mu) &\simeq 
  \eta^{8/\beta_0}\, I_1(\mu_0) \, G(\mu,\mu_0) \,, \qquad \eta= \frac{\alpha_s(\mu)}{\alpha_s(\mu_0)}
\end{align}
where we defined the RG-evolution function
\begin{align}
 G(\mu,\mu_0) &:= \exp\left[\frac{9}{8\pi^2} \, \int_{\mu_0}^{\mu} \frac{d\mu}{\mu} \, I_1(\mu) \right]
 \simeq \left(
   1 +  I_1(\mu_0) \, \frac{9}{4\pi} \, \frac{\eta-\eta^{8/\beta_0}}{(\beta_0-8)\, \alpha_s(\mu)}
  \right)^{-1} 
  \label{Gfunc} \,.
\end{align}
This coincides with \cite{Balzereit:1998id}, where the approximate RG flow of the
top Yukawa coupling has been derived (with $I_1(\mu) \simeq (y_t(\mu))^2$ 
and $\beta^{(1)}$ in \cite{Balzereit:1998id} is defined as $\beta^{(0)}/4$
in our convention.)
For the remaining invariants, using $y_c \gg y_s \gg y_{u,d}$ and $|\sin2\theta| \ll 1$, 
we have 
\begin{align}
\label{I2SM}
  \frac{dI_2}{d\ln\mu} &\simeq \frac{3}{8\pi^2} \, I_1 I_2 - \frac{4\alpha_s}{\pi} \, I_2  
  \qquad \Leftrightarrow \qquad \frac{d(I_2/I_1)}{d\ln\mu} \simeq - \frac{3}{4\pi^2} \, I_1 \left(
  \frac{I_2}{I_1} \right)
  \,,
\end{align}
and
\begin{align}
\label{I345SM}
  \frac{d\widehat I_3}{d\ln\mu} &\simeq \frac{15}{8\pi^2} \, I_1 \widehat I_3 - \frac{8\alpha_s}{\pi} \, \widehat I_3 
  & \Leftrightarrow \quad \frac{dx}{d\ln\mu} &\simeq - \frac{3}{8\pi^2} \, I_1 \, x \,,
  \cr 
  \frac{d\widehat I_4}{d\ln\mu} &\simeq \frac{9}{8\pi^2} \, I_1 \widehat I_4 - \frac{8\alpha_s}{\pi} \, \widehat I_4
  & \Leftrightarrow  \quad \frac{dy}{d\ln\mu} &\simeq \phantom{-} \frac{3}{8\pi^2} \, I_1 \, y \,,
  \cr 
  \frac{d\widetilde I_5}{d\ln\mu} &\simeq \frac{27}{8\pi^2} \, I_1 \widetilde I_5 - \frac{16\alpha_s}{\pi} \, \widetilde I_5
  & \Leftrightarrow  \quad \frac{dz}{d\ln\mu} &\simeq \phantom{-} \frac{3}{8\pi^2} \, I_1 \, z \,.
\end{align}
We see that once the RG-solution for $I_1(\mu)$ has been constructed, the RG equations for the 
remaining invariants can be easily solved by separation of variables.
Using the RG function $G(\mu,\mu_0)$ defined in (\ref{Gfunc}),
we have
\begin{align}
 I_2(\mu) &\simeq \eta^{8/\beta_0} I_2(\mu_0) \left[G(\mu,\mu_0)\right]^{1/3}
 \,,
\end{align}
and 
\begin{align}
 x(\mu) &\simeq  \left[G(\mu,\mu_0)\right]^{-1/3} x(\mu_0)
 \,,
 \quad 
 y(\mu) \simeq \left[G(\mu,\mu_0)\right]^{1/3} y(\mu_0)
 \,, 
 \quad 
 z(\mu)  \simeq \left[G(\mu,\mu_0)\right]^{1/3} z(\mu_0)
 \,. 
 \label{xyz}
\end{align}

\subsubsection{Numerical Illustration}

\begin{figure}[t!pbh]
\begin{center}
\includegraphics[width=0.45\textwidth]{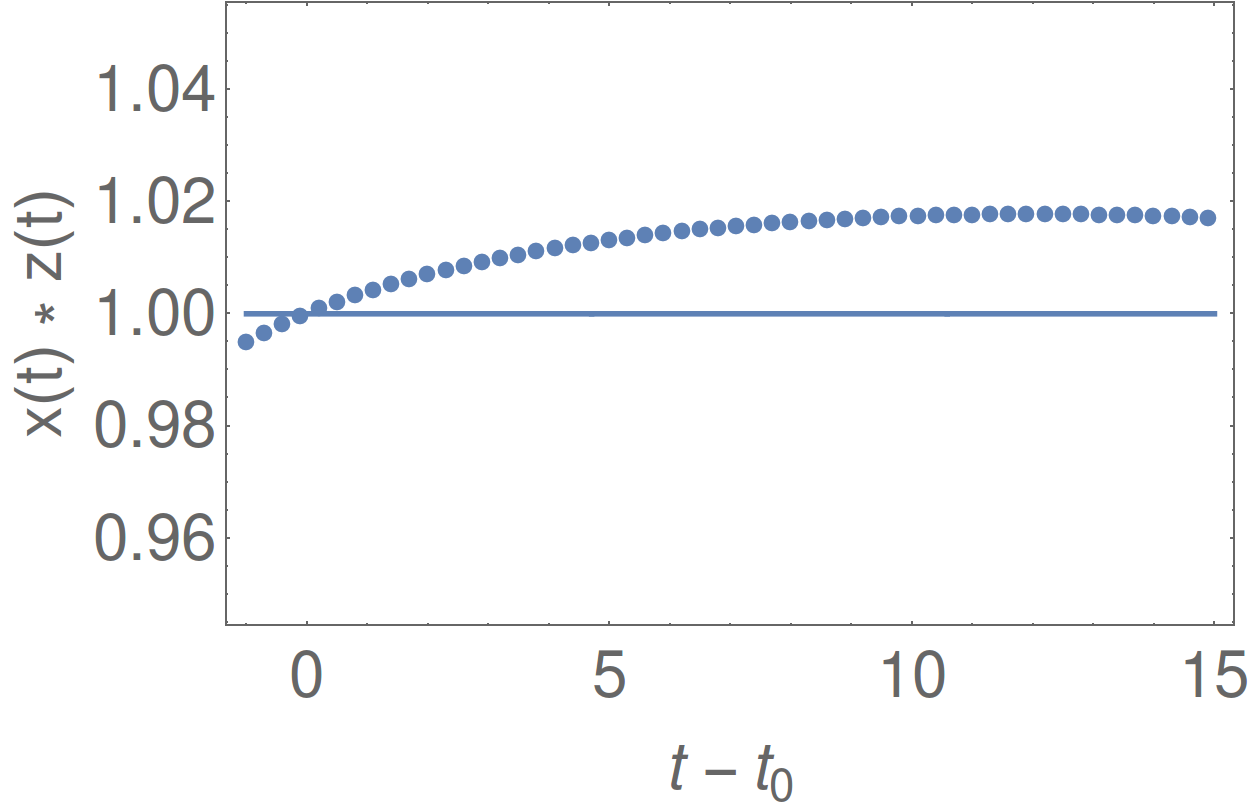} \qquad 
\includegraphics[width=0.45\textwidth]{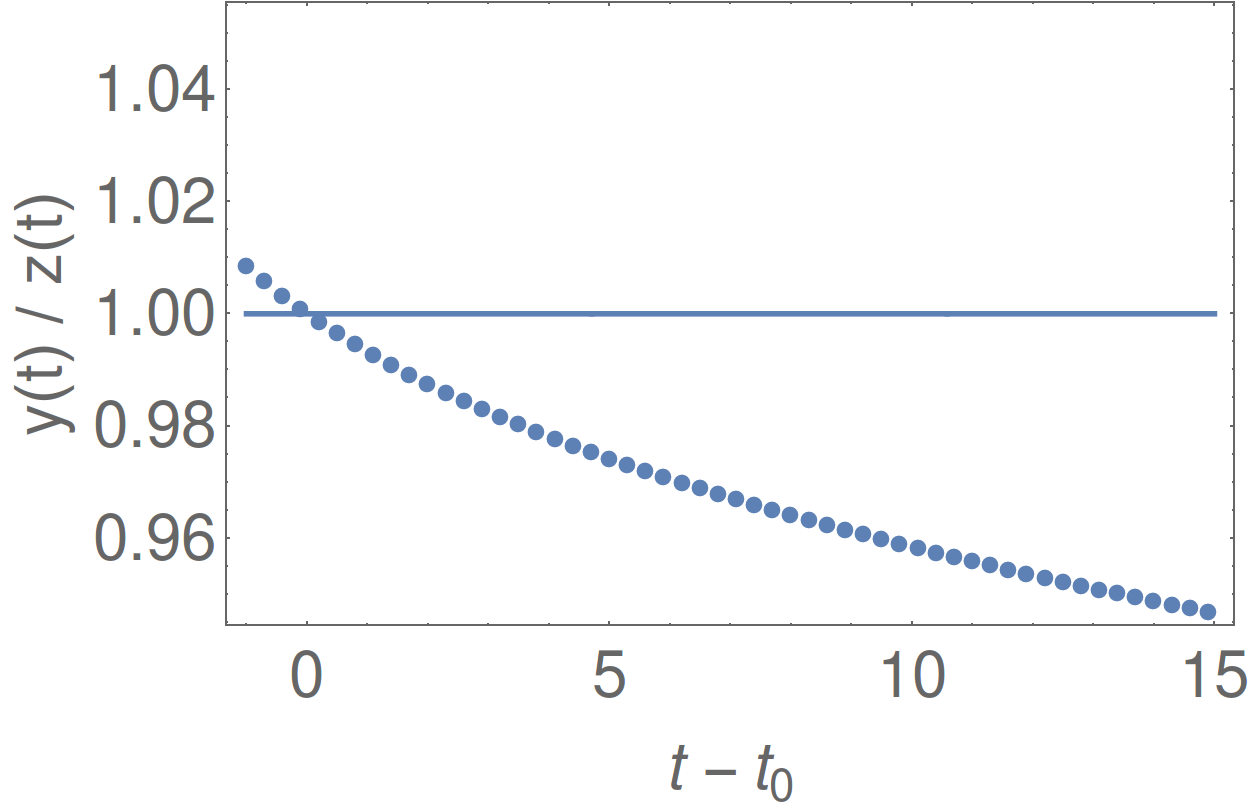} 
\end{center}
\caption{\label{fig1:num2G} 
Comparison of numerical [dots] and approximate analytical (\ref{xyz})
[solid line]
solution for combinations of flavour invariants, $x(t) z(t)$ and 
$y(t)/z(t)$, 
normalized to the values at $t=\ln\mu/\mu_0 \to t_0= 0$.
The following starting values have been used:
$I_1(t_0)=1.0$, 
$I_2(t_0)=0.1$,
$x(t_0)=0.005$,
$y(t_0)=0.048$,
$z(t_0)=0.105$.
}
\end{figure}

In Fig.~\ref{fig1:num2G} we provide 
illustrations for the one-loop
RG flow of the combinations of flavour invariants $x,y,z$ in the SM,
and compare the exact numerical solutions with the approximation
in (\ref{xyz}). We observe that ---
for the chosen numerical starting values --- even for
values as large as $t=\ln \mu/\mu_0 =15$, the differences between
the exact and approximate solutions are always below 5\%.

\begin{figure}[t!pbh]
\begin{center}
\begin{tabular}{ccc}
\includegraphics[width=0.45\textwidth]{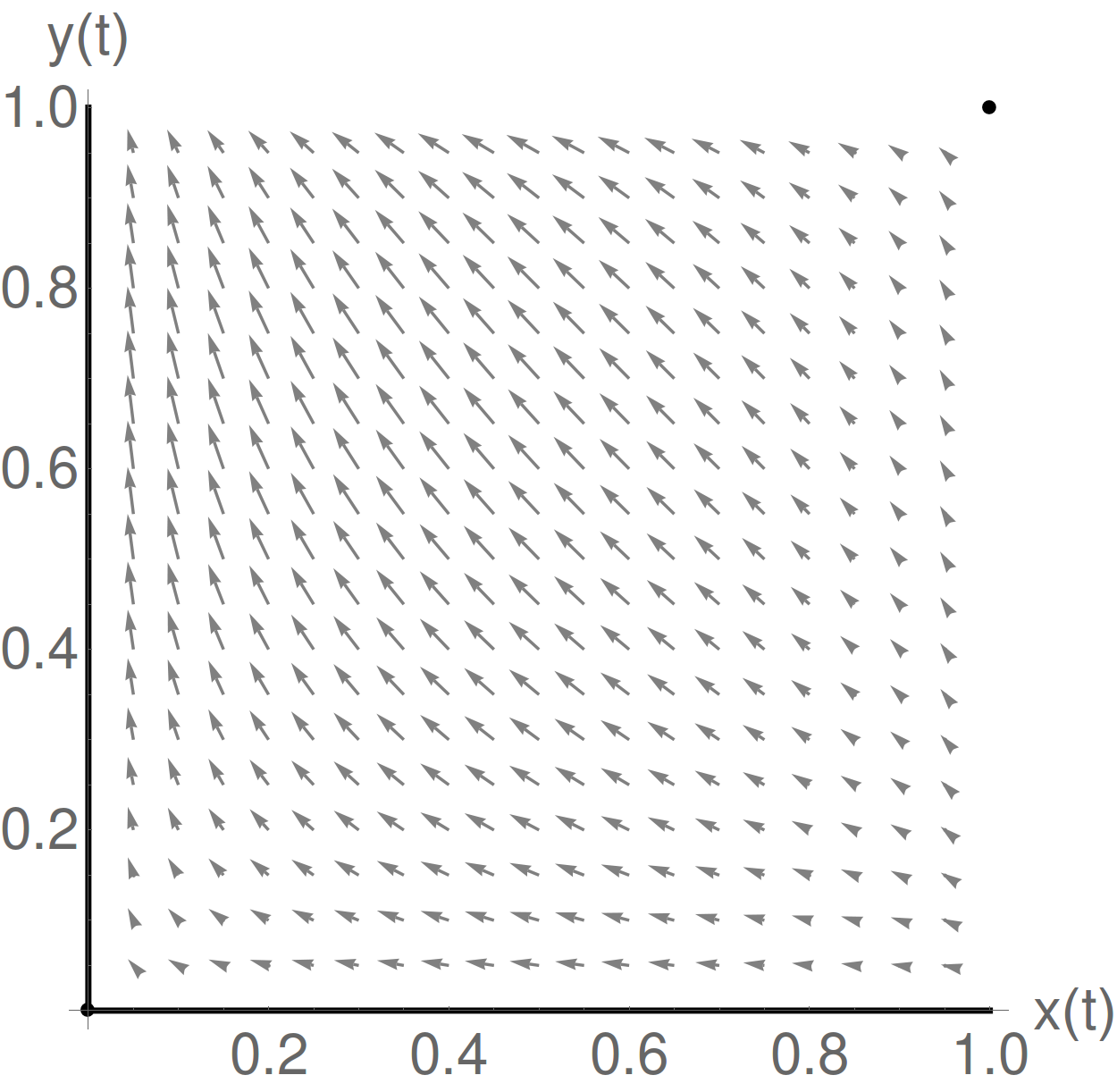} & \hspace{3em} &
\includegraphics[width=0.45\textwidth]{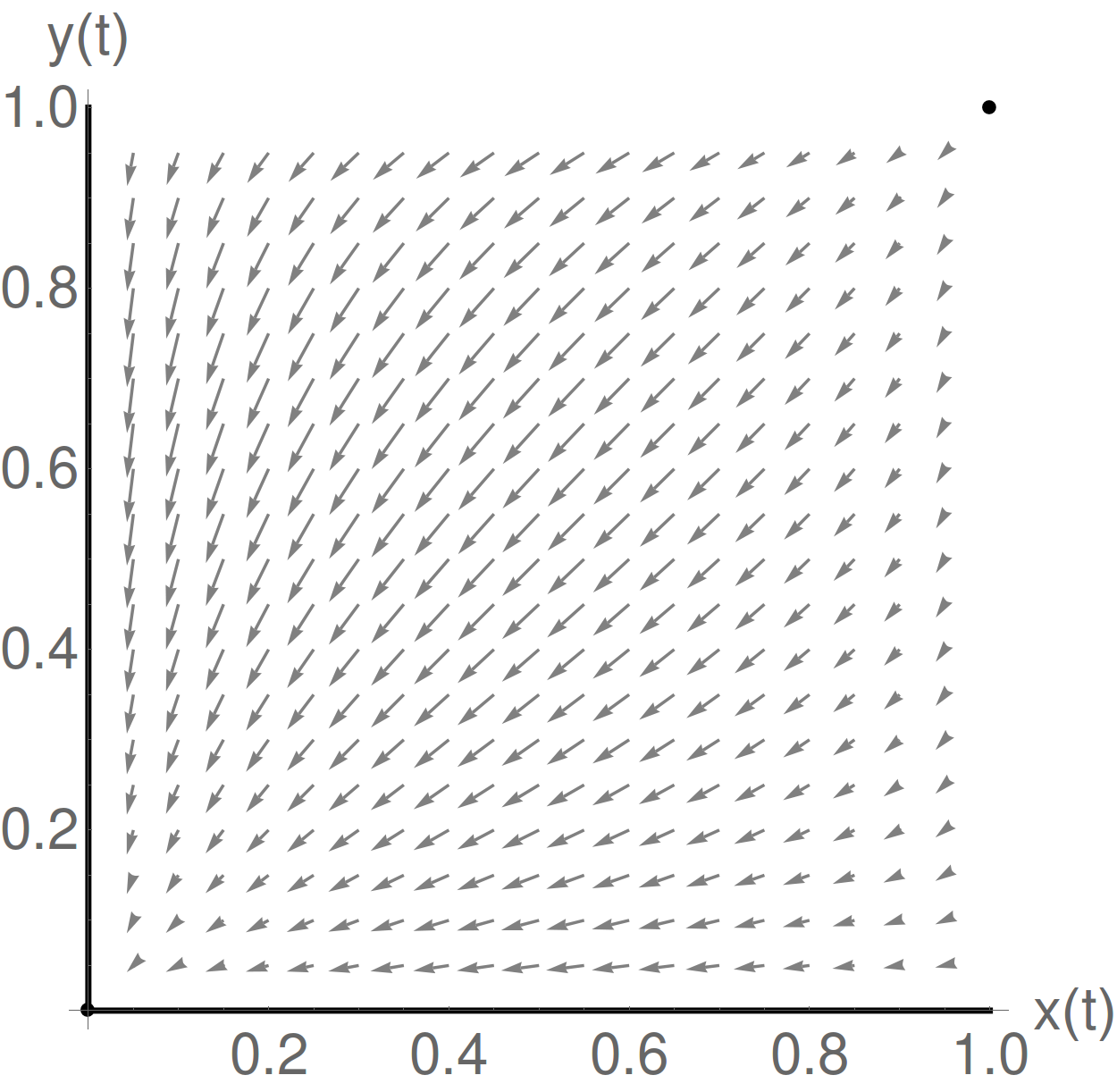}
\\
\includegraphics[width=0.45\textwidth]{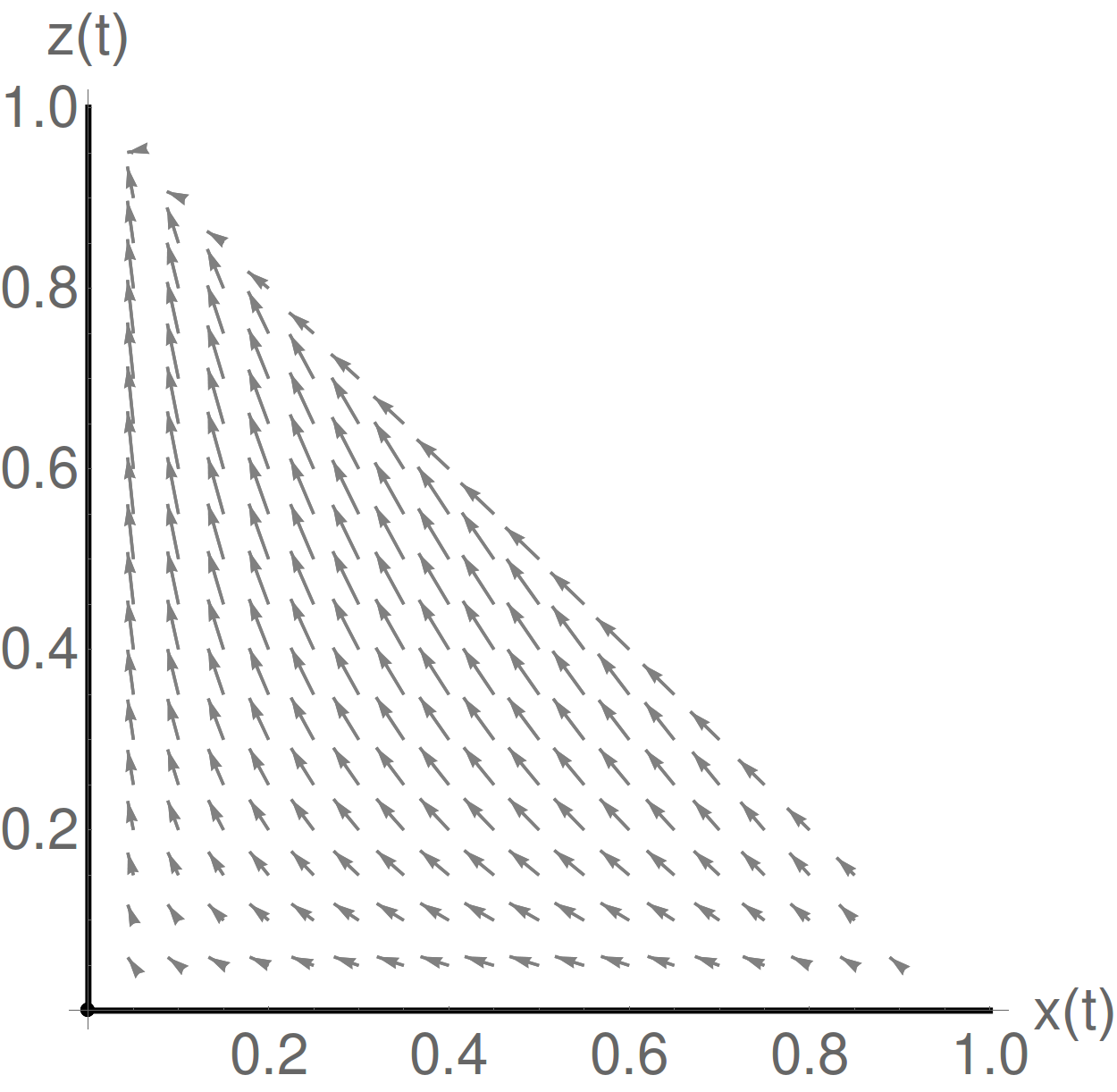} & \hspace{3em} &
\includegraphics[width=0.45\textwidth]{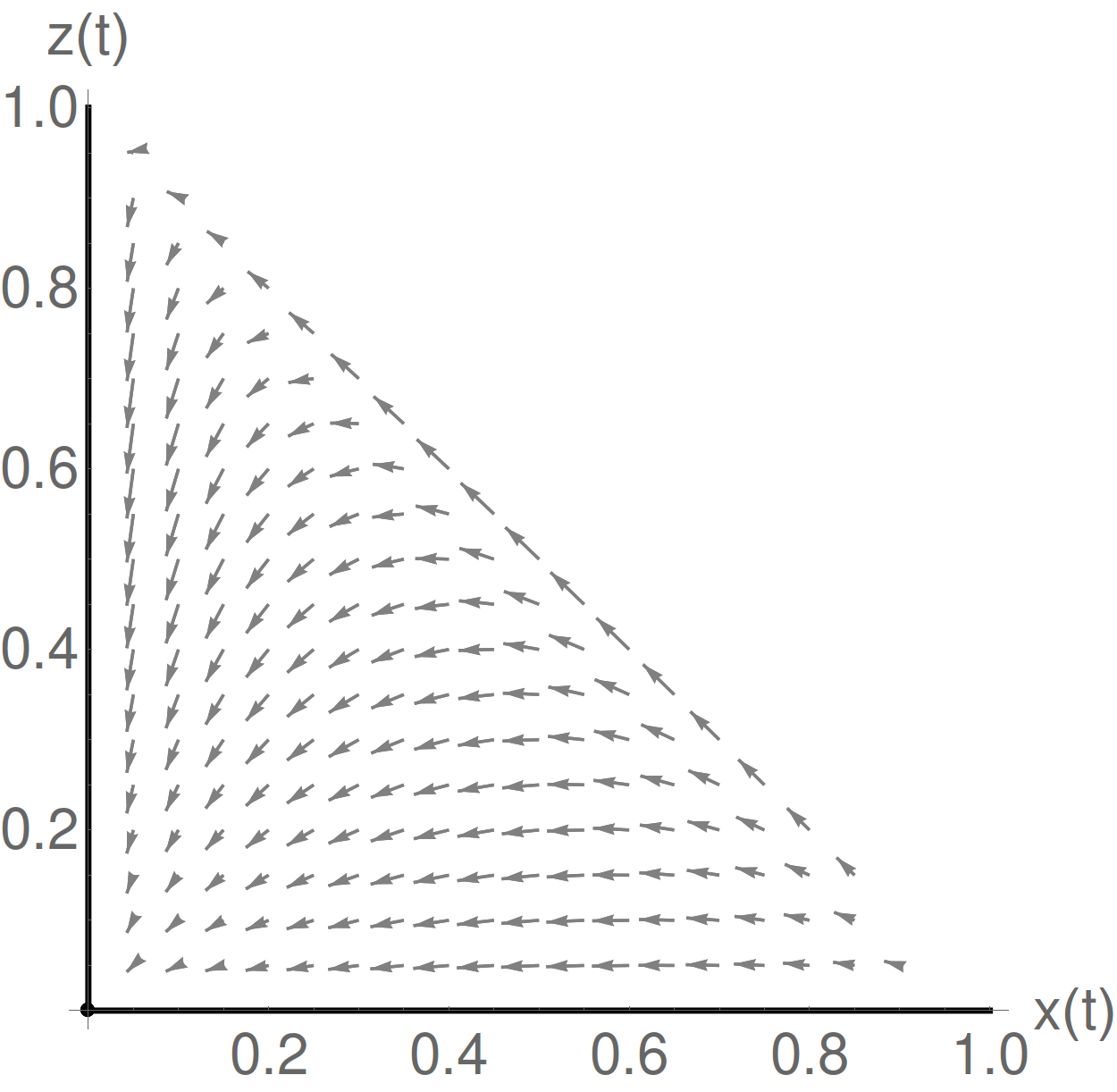}
\\
\includegraphics[width=0.45\textwidth]{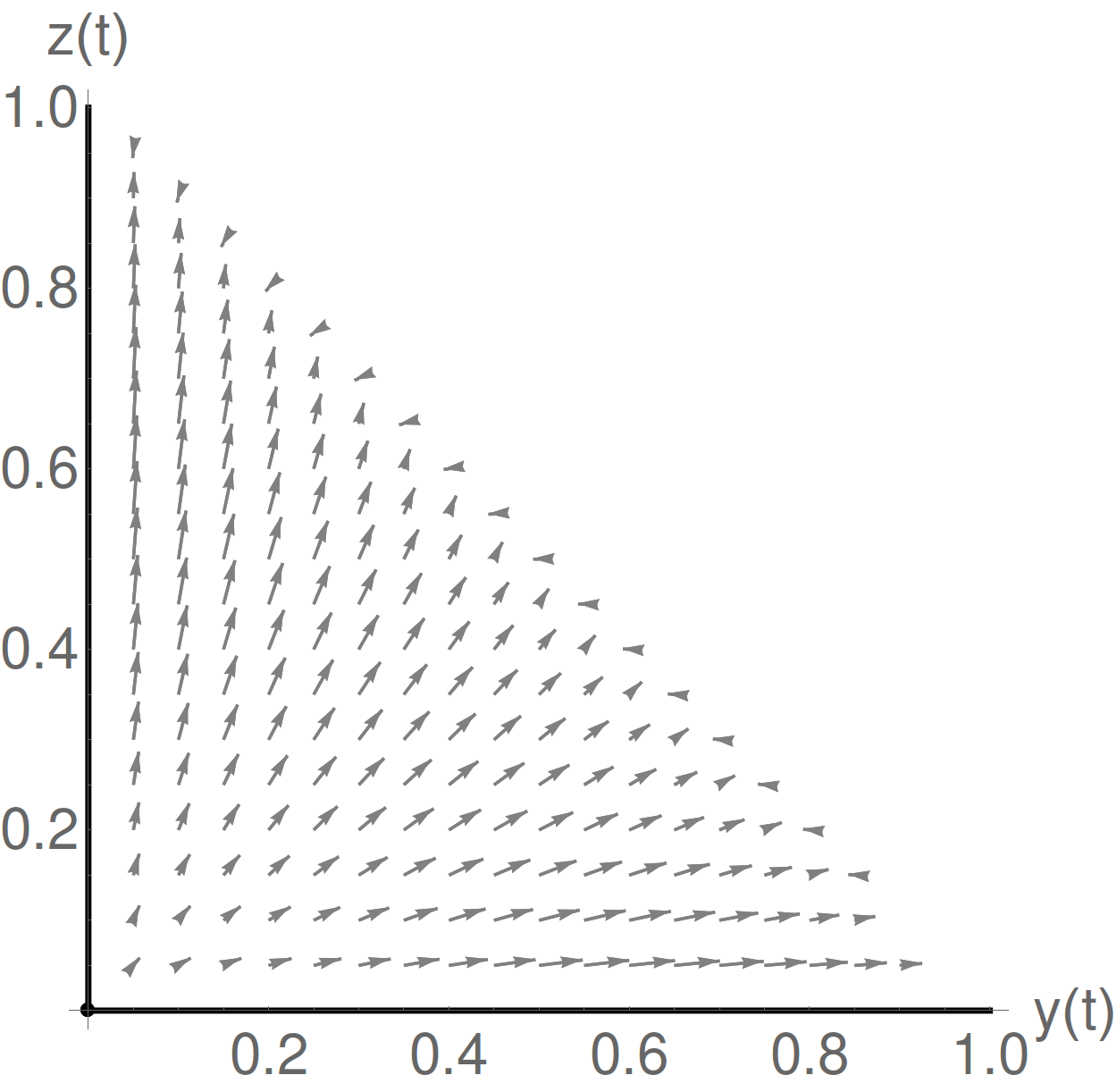} & \hspace{3em} &
\includegraphics[width=0.45\textwidth]{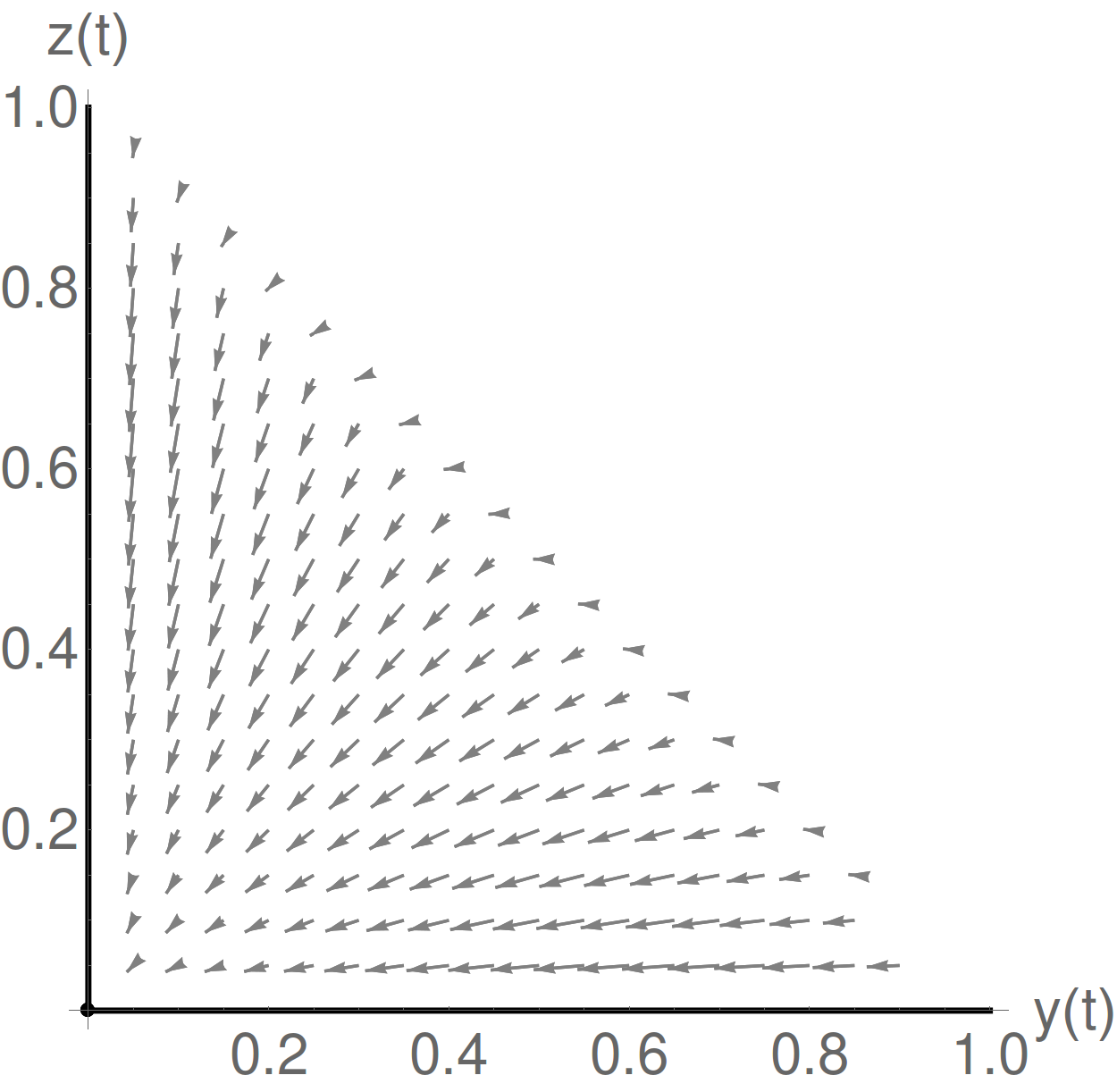}
\end{tabular}
\end{center}
\caption{\label{fig2:num2G} Numerical illustration of the  RG flow 
at the ``phase-space boundaries'' for flavour invariants
in the SM (2G, one-loop accuracy, neglecting electroweak
gauge couplings). 
Top: $z\equiv 0$; center: $y \equiv 0$; bottom $x \equiv 0$. 
Each arrow indicates the RG flow from $t_0=0$ to $t=5$; the 
starting values are again chosen as 
$I_1(\mu_0)=1.0$, 
$I_2(\mu_0)=0.1$,
$\alpha_s(\mu_0)=0.2$.
The plots on the left (right) are generated assuming 
$I_5(\mu_0)= \mp \sqrt{I_3 I_4 - \widetilde I_5}(\mu_0)$.
}
\end{figure}

In Fig.~\ref{fig2:num2G} we illustrate the RG flow for the 
boundaries of the ``phase-space'' of flavour invariants, defined 
by $z\equiv 0$, $y\equiv 0$, $x\equiv 0$, respectively, see the discussion
in Section~\ref{subsec:disc}. Again,
we have chosen a hierarchical scenario with $I_2(\mu_0) \ll I_1(\mu_0)$.
We observe that 
\begin{itemize}
  \item The relation $y(t)/z(t) \simeq \text{const.}$ holds 
    on the whole plane $x=0$, which is in line with our derivation 
    of (\ref{yz}) which only required $I_1 \gg I_2$.
  \item In contrast, $x(t) y(t) \simeq \text{const.}$ only 
    holds in the vicinity of $x \sim y \sim 0$ (where $y_u \ll y_c$ and $y_d \ll y_s$)
    and for $\theta$ near zero (which requires the solution 
    with $I_5(\mu_0)= -\sqrt{I_3 I_4 - \widetilde I_5}(\mu_0)I_5$ shown on the 
    left-hand side). 
  \item The same is true for $x(t) z(t) \simeq \text{const.}$.
\end{itemize}

\clearpage

\section{Three Quark Generations}

\def\adj{{\rm adj}\,}

\label{sec:3G}

For three quark generations in the SM, the flavour symmetry group to consider now is
\begin{align}
  {\cal G}_{\rm quark} &= U(3)^3/U(1)_B \sim SU(3)_{Q_L} \times U(3)_{U_R} \times U(3)_{D_R} \,.
\end{align}
The corresponding Yukawa matrices again transform as bi-doublets under a change of flavour basis, 
$$
 Y_U \to V_{Q_L} Y_U V_{U_R}^\dagger \,, \qquad Y_D \to V_{Q_L} Y_D V_{D_R}^\dagger \,.
$$
In a particular flavour basis, they are given by
\begin{align}
  & Y_U = \left( \begin{array}{ccc} y_u & 0 & 0 \\
  0 & y_c & 0 \\
  0 & 0 & y_t
  \end{array} \right) \,,  \qquad 
  Y_D = V_{\rm CKM} \left( \begin{array}{ccc} y_d & 0 & 0 \\
  0 & y_s & 0 \\
  0 & 0 & y_b
  \end{array} \right) \,.
\end{align}
In the subsequent analysis, it turns out to be more convenient to discuss 
the flavour invariants as a function of the CKM elements $V_{ij}$ without choosing a particular
parametrization in terms of mixing angles 
which would directly reflect the unitarity of the CKM matrix $V_{\rm CKM}$.

\subsection{Flavour Invariants}

\label{sec:3Ginvariants}

As discussed in \cite{Jenkins:2009dy}, the SM quark sector in the 3G case 
can be described in terms of $10+1$ polynomially independent invariants,
which determine 6 Yukawa eigenvalues, 3 mixing angles and the cosine and
sine of the CP-violating phase (in a given parametrization of the CKM matrix). 
With a similar procedure as in the 2G case,
we will now explicitly construct a convenient set for these 11 invariants
from the non-negative hermitian matrices,
$$
  U \equiv Y_U Y_U^\dagger\,, \qquad D \equiv Y_D Y_D^\dagger \,,
$$
which now transform under the $SU(3)_{Q_L}$ flavour symmetry. 
For later use we also define the adjoint matrices, satisfying
$$
U \, \adj U = \det U \quad \mbox{etc.}
$$
With this, we can easily construct a complete set of polynomially
independent positive semi-definite invariants. For the unmixed
invariants, we define
\begin{align}
I_1 & \equiv \tr \left(U\right) \geq 0 \,, &&  I_2 \equiv \tr \left(D\right) \geq 0 \,, 
\cr 
\widehat I_3 & \equiv  \tr \left(\adj U \right) \geq 0 \,,
 &&
\widehat I_4 \equiv  \tr \left(\adj D \right) \geq 0 \,,
\cr 
\widehat I_6 &\equiv \det \left(U\right) \geq 0 \,, &&
\widehat I_8 \equiv \det \left(D\right) \geq 0 \,,  \label{3Gpos1}
\,.
\end{align}
These determine the six singular values of the Yukawa matrices.
The CKM elements are then determined by mixed invariants which
we define in a similar way. CP-even invariants can be chosen as
\begin{align}
\widehat I_5  \equiv \tr \left(UD\right) \geq 0 \,, \qquad 
\widehat I_7  \equiv \tr \left(D \, \adj U \right) \geq 0  \,, \qquad 
  \widehat I_9 \equiv \tr \left( U \, \adj D\right) \geq 0 \,, 
  \label{3Gpos2}
\end{align}
and 
\begin{align}
 \widehat I_{10} & \equiv \tr \left(\adj (UD) \right)\geq 0 \,, 
\label{3Gpos3}
\end{align}
According to the discussion in \cite{Jenkins:2009dy},
there is an eleventh, CP-odd, invariant that cannot be expressed
as a polynomial of the other ten invariants, as defined above. It is related to the 
Jarlskog determinant \cite{Jarlskog:1985ht} and can be chosen as
\begin{align}
 I_{11}^- & 
 = - \frac{3i}{8} \, \det\left[U,D\right] \,.
\end{align}
Explicit expression in terms of Yukawa couplings and
CKM elements can be found in Appendix~\ref{translate}.

\paragraph{Octet Matrices and Octet Invariants:}

As in the 2G case, we can also construct 
basic flavour matrices as octet representations of
the flavour group factor $SU(3)_{Q_L}$. First, there are two polynomially 
independent octet matrices that are quadratic in the Yukawas, namely the traceless part
of the matrices $U$ and $D$ (defined analogously to the 2G case),
\begin{align}
& U_8 \equiv U- \frac13 \, \tr[U] \, \mathds{1} \,,
 \qquad 
 D_8 \equiv U- \frac13 \, \tr[D] \, \mathds{1} \,.
 \end{align}
 For terms quartic in the Yukawas, we may define the octet part of the 
 adjoint matrices, 
\begin{align*}
&   \adj U_8 \equiv \adj U - \frac13 \, \tr[\adj U]\, \mathds{1}\,,
\qquad 
   \adj D_8 \equiv \adj D - \frac13 \, \tr[\adj D] \, \mathds{1} \,,
 \end{align*}
 together with 
 \begin{align}
 & S_8 \equiv  \frac12 \left\{ U,D \right\} - \frac13 \, \tr[U D] \, \mathds{1} \,,
 \qquad 
 A_8 \equiv  \frac{i}{2} \left[U,D\right] \,.
 \end{align}
Similarly, we define 
\begin{align}
& A_U \equiv  \frac{i}2 \left[\adj U, D\right] 
\,, \qquad
A_D \equiv  \frac{i}2 \left[\adj D , U \right]  \,.
\end{align}

\begin{table}

\begin{center}
  \renewcommand{\arraystretch}{2} 
  \begin{tabular}{|c||ccccc|ccc|}
  \hline 
  $\tr \left[  \ \right]$  &  $U_8$ & $D_8$ & $\adj U_8$ & $\adj D_8$
                    & $ S_8$ & $A_8$ & $A_U$ & $A_D$
                    \\
   \hline \hline 
   $U_8$ & 
   $I_3$ & $I_5$ & $I_6$ & $I_9$ & $\widetilde I_7$ & 0 & 0 & 0
   \\
   $D_8$ &
     +   & $I_4$ & $I_7$ & $I_8$ & $\widetilde I_9$ & 0 & 0 & 0
\\ 
\hline 
   $\adj U_8$  & 
     +   &     + & $P_1$ & $P_2$ & $ P_3 $ & 0 & 0 & 0
   \\
   $\adj D_8$  &
     +   &     + &                + 
         & $P_4$ & $P_5$ & 0 & 0 & 0
   \\
   $S_8$  & 
   + & + & + & + 
   & $I_{10}$ & 0 & 0 & 0
   \\
   \hline 
   $A_8$ & 0 & 0 & 0 & 0 & 0 
     & $\widetilde I_{10}$ 
     & $Q_1$ 
     & $Q_2$  
   \\
   $A_U$ & 0 & 0 & 0 & 0 & 0 & + 
     & $Q_3$ 
     & $Q_4$ 
   \\
   $A_D$ & 0 & 0 & 0 & 0 & 0 & + & + 
     & $Q_5$ 
   \\
   \hline 
  \end{tabular}
  \end{center}
  \caption{\label{metric3G} 
  Traces of basic octet matrices constructed from 
  $Y_U Y_U^\dagger$ and $Y_D Y_D^\dagger$ as defined in the text. }
\end{table}  

For generic Yukawa entries, the eight hermitian matrices 
as defined above provide a basis for octet matrices in $SU(3)_{Q_L}$. 
The symmetric (but non-orthogonal) 
metric defined by the traces of matrix products contains 
flavour invariants for the $3\times 3$ case. 
It is summarized in Table~\ref{metric3G}.
Here, the unhatted invariants are related to the hatted ones via
\begin{align}
  I_3 \equiv \tr [U_8^2] & = \frac23 I_1^2 - 2 \widehat I_3 
  \,, &&
  I_4 \equiv \tr [D_8^2]  = \frac23 I_2^2 - 2 \widehat I_4 
  \,,
  \cr 
  I_6 \equiv \tr[U_8 \, \adj U_8] & = 3 \widehat I_6 - \frac{I_1 \widehat I_3}{3}
  \,, &&
  I_8 \equiv \tr[D_8 \, \adj D_8]  = 3 \widehat I_8 - \frac{I_2 \widehat I_4}{3}
  \,,
\end{align}
and 
\begin{align}
  I_5 &\equiv \tr[U_8 D_8] = \widehat I_5 - \frac{ I_1 I_2}{3} \,,
\end{align}  
and 
\begin{align}
   I_7 &\equiv \tr[D_8 \,\adj U_8 ] = \widehat I_7 - \frac{I_2 \widehat I_3}{3}  \,,
  \qquad 
  I_9 \equiv  \tr[U_8 \, \adj D_8 ] = \widehat I_9 - \frac{I_1 \widehat I_4}{3} 
\,,
\end{align}
and 
\begin{align}
  \widetilde I_7 &\equiv \tr[U_8  S_8] = \widehat I_7 - I_2 \widehat I_3 + \frac{2 I_1 \widehat I_5}{3} \,,
  \cr  
  \widetilde I_9 &\equiv  \tr[D_8 S_8] = \widehat I_9 - I_1 \widehat I_4 + \frac{2 I_2 \widehat I_5}{3}
\,,
\end{align}
Finally, one has
\begin{align}
 I_{10} & \equiv \tr[ S_8  S_8] = 
 \frac{\widehat I_3 \widehat I_4 -\widehat I_{10}}{2} 
 + \frac{(\widehat I_5 - I_1 I_2) \, \widehat I_5}{6} 
 + \frac{I_1 \widetilde I_9 + I_2\widetilde I_7}{2}
 \,.
 \end{align}
Further polynomially dependent invariants that appear in Table~\ref{metric3G}
are given by
\begin{align}
\widetilde I_{10} & \equiv  \tr[A_8^2] = I_{10} + 2 \widehat I_{10} - \frac{2 \widehat I_5^2}{3} \,, 
\end{align}
and 
\begin{align}
P_1 \equiv \tr[\adj U_8 \, \adj U_8] & = - 2 I_1 \widehat I_6 + \frac{2 \widehat I_3^2}{3}  \,,
\qquad 
P_4 \equiv \tr[\adj D_8 \, \adj D_8]  = - 2 I_2 \widehat I_8 + \frac{2 \widehat I_4^2}{3}  \,,
\end{align}
and 
\begin{align}
P_2 \equiv \tr[\adj U_8 \, \adj D_8] & = \widehat I_{10} - \frac{\widehat I_3 \widehat I_4}{3} \,,
\end{align}
and
\begin{align}
P_3 \equiv \tr[\adj U_8 \, S_8] & = I_2 \widehat I_6 - \frac{\widehat I_3 \widehat I_5}{3} \,,
\qquad 
P_5 \equiv \tr[\adj D_8 \, S_8]  = I_1 \widehat I_8 - \frac{\widehat I_4 \widehat I_5}{3} \,.
\end{align}
Furthermore,
\begin{align} 
Q_1 &\equiv \tr[ A_8 \, A_U] = 
- \frac{I_1 \, (\widehat I_{10} - \widehat I_3 \widehat I_4)}{2}
+ \frac{3 I_4 \widehat I_6}{4} - \frac{\widehat I_5 \widehat I_7}{2} 
-\frac{\widehat I_3 \widehat I_9}{2} \,,
\cr 
Q_2&\equiv \tr[ A_8 \, A_D] = 
+ \frac{I_2 \, (\widehat I_{10} - \widehat I_3 \widehat I_4)}{2}
- \frac{3 I_3 \widehat I_8}{4} + \frac{\widehat I_5 \widehat I_9}{2} 
+\frac{\widehat I_4 \widehat I_7}{2}
\,,
\end{align}
and
\begin{align} 
Q_3 &\equiv \tr[ A_U \, A_U] = 
\frac{\widehat I_3 \, (\widehat I_{10} - \widehat I_3 \widehat I_4)}{2}
+ \frac{\widehat I_6 \, (3 I_9 - I_1 I_4 + I_2 I_5)}{2} 
+ \frac{(I_2 \widehat I_3 - \widehat I_7) \, \widehat I_7}{2}
\,,
\cr 
Q_5 &\equiv \tr[ A_D \, A_D] = 
\frac{\widehat I_4 \, (\widehat I_{10} - \widehat I_3 \widehat I_4)}{2}
+ \frac{\widehat I_8 \, (3 I_7 - I_2 I_3 + I_1 I_5)}{2} 
+ \frac{(I_1 \widehat I_4 - \widehat I_9) \, \widehat I_9}{2}
\end{align}
and 
\begin{align}
Q_4 \equiv \tr[A_U \,A_D] & =
 \frac{(\widehat I_5 - I_1 I_2)  (\widehat I_{10} - \widehat I_3 \widehat I_4)}{4}
 + \frac{(\widehat I_7- I_2 \widehat I_3)(\widehat I_9 - I_1 \widehat I_4)}{4}
 \cr & \quad 
 - \frac{(\widehat I_6 - I_1 \widehat I_3)(\widehat I_8 - I_2 \widehat I_4)}{4}
 - 2 \widehat I_6 \widehat I_8 \,.
\end{align}
In order to project onto the eight basis matrices 
one needs the inverse of the metric in Table~\ref{metric3G}. 
The explicit result is rather lengthy and not very instructive,
and we therefore refrain from quoting it here. We checked however
that the metric is not singular for generic Yukawa entries.

\subsection{One-Loop RG equations}

In the 3G case, again,
any flavour matrix that arises as a flavour-covariant product of SM Yukawa matrices 
$Y_U$ and $Y_D$ can be written as a linear combination of a finite set of basic matrices
(constructed from $Y_U$ and $Y_D$) with coefficients given as
\emph{polynomials} of a finite number of flavour invariants (as a corollary to
the discussion in \cite{Jenkins:2009dy}).
The most general form of the RG equations then can be written as
\begin{align}
 \frac{d Y_U}{d\ln\mu} &= 
\left( a_0 \, \mathds{1} 
+ a_1 \, U_8 + a_2 \, D_8 
+ a_3 \, \adj U_8 + a_4 \, \adj D_8 
+ a_5 \, S_8 + i a_6 \, A_8 \right. \cr 
& \qquad \left. + i a_7 \, A_U + i a_8 A_D \right) Y_U(\mu) \,,
\cr
 \frac{d Y_D}{d\ln\mu} &= 
\left( b_0 \, \mathds{1} 
+ b_1 \, D_8 + b_2 \, U_8 
+ b_3 \, \adj D_8 + b_4 \, \adj U_8 
+ b_5 \, S_8 -i b_6 \, A_8 \right. \cr 
& \qquad \left.- i b_7 \, A_D - i b_8 A_U \right) Y_D(\mu)
\,,
\label{eq:3Grge-generic}
\end{align} 
As compared to the 2G case, the expressions for the RG equations
of the flavour invariants derived from this general parametrization 
become rather lengthy.
(The explicit structure of the two-loop expressions can be found in (\ref{3G-2loop})
in the appendix.)

\subsubsection{General Form}

If we restrict ourselves to the RG equations at one-loop accuracy.
we can write
\begin{align}
 \frac{d Y_U}{d\ln\mu} &= 
\left( a_0 \, \mathds{1} + a_1 \, U_8 + a_2 \, D_8 
+ \ldots \right) Y_U(\mu) \,,
\cr
 \frac{d Y_D}{d\ln\mu} &= 
\left( b_0 \, \mathds{1} + b_1 \, D_8 + b_2 \, U_8 
+ \ldots \right) Y_D(\mu)
\,,
\label{3G-1loop}
\end{align}
where the coefficients $a_0,b_0$ are first-order polynomials 
of flavour invariants, and $a_{1,2},b_{1,2}$ are constant,
see again Table~\ref{tab:aibi}.
The RG equations for the quadratic invariants then take the same
form as in the 2G-case,
\begin{align}
  \frac{dI_{1}}{dt} &\simeq 2 a_{0} \, I_{1} + 2a_{1} \, I_{3} + 2a_{2} \, I_{5}  \,,
  \qquad 
  \frac{dI_{2}}{dt} \simeq  2 b_{0} \, I_{2} + 2b_{1} \, I_{4} + 2b_{2} \, I_{5}  \,.
\end{align}
We remind the reader of the difference between the hatted and unhatted
invariants, as defined in Section~\ref{sec:3Ginvariants}.
For the remaining unmixed invariants, 
we also find simple expressions
\begin{align}
  \frac{d\widehat I_{3}}{dt} &\simeq 
   4 a_{0} \, \widehat I_{3}
   - 2 a_1 \, I_6
   - 2 a_2 \, I_7
  \,,
  \qquad  
  \frac{d\widehat I_{4}}{dt} \simeq
   4 b_{0} \, \widehat I_{4}
   - 2 b_1 \, I_8
   - 2 b_2 \, I_9
\,,
 \end{align} 
and
\begin{align}
 \frac{d \widehat I_6}{dt} &=
 6 a_0 \, \widehat I_6 \,, \qquad 
 \frac{d \widehat I_8}{dt} =
 6 b_0 \, \widehat I_8 \,.
\end{align}
Notice that the last two relations --- with our convention in (\ref{eq:3Grge-generic})
where the coefficients $a_{i>0},b_{i>0}$ always multiply \emph{traceless} matrices
--- are exact.
The one-loop RG equations for the mixed invariants are determined as
\begin{align}
 \frac{d \widehat I_5}{dt} & \simeq
 (2 a_0 + 2 b_0) \, \widehat I_5 
+ (2 a_1 + 2 b_2) 
\, \widetilde I_7
+ (2 a_2 + 2 b_1) 
\, \widetilde I_9
\,,
 \end{align}
together with 
\begin{align}
 \frac{d \widehat I_7}{dt} &=
 (4 a_0 + 2 b_0) \, \widehat I_7 
 + (2 a_1 - 2 b_2) \left( 
 \frac{I_1 \widehat I_7}{3} - I_2 \widehat I_6 \right)
 \cr & \quad 
 - (2 a_2 - 2 b_1) \left(
 \widehat I_{10} - \widehat I_3 \widehat I_4 + \frac{2 I_2 \widehat I_7}{3}
 \right)
 \,, \\
 \frac{d \widehat I_9}{dt} &=
 (4 b_0 + 2 a_0) \, \widehat I_9 
 + (2 b_1 - 2 a_2) \left( 
 \frac{I_2 \widehat I_9}{3} - I_1 \widehat I_8 \right)
 \cr & \quad 
- (2 b_2 - 2 a_1) \left(
 \widehat I_{10} - \widehat I_3 \widehat I_4 + \frac{2 I_1 \widehat I_9}{3}
 \right)
 \,,
 \end{align}
and 
 \begin{align}
\frac{d\widehat I_{10}}{dt} &= 
(4 a_0 + 4 b_0) \, \widehat I_{10} 
+ (2 a_1 + 2 b_2) \left( \frac{I_1 \widehat I_{10}}{3} - \widehat I_4 \widehat I_6 \right) 
+ (2 a_2 + 2 b_1) \left( \frac{I_2 \widehat I_{10}}{3} - \widehat I_3 \widehat I_8 \right)
\,.
 \end{align}
The RG equations for the Jarlskog invariant is simple, and to one-loop accuracy reads
 \begin{align}
  \frac{dI_{11}^-}{dt} &=  \left(6 a_0 + 6 b_0 + 2 a_1 \, I_1 + 2 b_1 \,  I_2 \right) I_{11}^- \,. 
 \end{align}

\subsubsection{Exploiting Flavour Hierarchies}

The RG equations again simplify when one exploits flavour hierarchies
in the SM Yukawa matrices, which are also applicable to 
MFV extensions of the SM. 
For concreteness, we relate the scaling of the quark Yukawa couplings
to the Wolfenstein parameter $\lambda$ in the CKM matrix,
as it can be realized in Froggatt-Nielsen models \cite{Froggatt:1978nt}
(see also \cite{Feldmann:2006jk,Feldmann:2009dc}), assuming
\begin{align}
&  V_{12} \sim \lambda \,, \qquad 
  V_{23} \sim \lambda^2 \,, \qquad 
  V_{13} \sim \lambda^3 \,, \
\end{align}
and
\begin{align}
 & y_t \sim \lambda^0 \,, \quad  y_b \sim \lambda^2 \,,
 \quad y_c \sim \lambda^3 \,, \quad 
 y_s \sim \lambda^6 \,, \quad y_{u,d} \sim \lambda^8 \,.
\end{align}
Defining $\epsilon=\lambda^2$, the individual invariants scale as 
(see Appendix~\ref{translate})
\begin{align}
& I_1 \simeq  y_t^2 \sim \epsilon^0 \,,  \qquad 
\widehat I_3 \simeq y_t^2 y_c^2 \sim \epsilon^3 \,, \qquad 
\widehat I_6 = y_t^2 y_c^2 y_u^2 \sim \epsilon^{11} \,,
\cr 
& I_2 \simeq y_b^2 \sim \epsilon^2 \,,  \qquad 
\widehat I_4 \simeq y_b^2 y_s^2 \sim \epsilon^8 \,, \qquad 
\widehat I_8 = y_b^2 y_s^2 y_d^2 \sim \epsilon^{16} \,,
\end{align}
and
\begin{align}
 \widehat I_5  \simeq y_t^2 y_b^2 |V_{tb}|^2  = I_1 I_2 + {\cal O}(\epsilon^4) \,,
 \qquad &
 \widehat I_7  \simeq y_t^2 y_b^2 y_c^2 |V_{ub}|^2 \sim \epsilon^8 \,,
 \cr 
 \widehat I_{10} \simeq y_t^2 y_b^2 y_c^2 y_s^2 |V_{ud}|^2 = \widehat I_3 \widehat I_4 
 + {\cal O}(\epsilon^{12}) \,,
 \qquad &
 \widehat I_9 \simeq y_t^2 y_b^2 y_s^2 |V_{td}|^2 \sim \epsilon^{11} \,, 
\end{align}
and 
\begin{align}
I_{11}^- & \sim \epsilon^{16} \,.
\end{align}
The leading terms in the (one-loop) RG equations are then identified as\footnote{ 
We do not include the power-counting for the gauge-coupling constants here
as has been advocated in \cite{JuarezWysozka:2002kx}.}
\begin{align}
\frac{d I_1}{dt} & \simeq \left( 2 a_0  + \frac{4 a_1}{3} \, I_1 \right) I_1 \,, 
&&
\frac{dI_2}{dt}  \simeq \left( 2 b_0  + \frac{4 b_2}{3} \, I_1 \right) I_2 \,,
\cr 
\frac{d \widehat I_3}{dt} & \simeq \left( 4 a_0  + \frac{2 a_1}{3} \, I_1 \right) \widehat I_3 \,,
&&
\frac{d\widehat I_4}{dt}  \simeq \left( 4 b_0  + \frac{2 b_2}{3} \, I_1 \right) \widehat I_4 \,,
\end{align}
and
\begin{align}
\frac{d (\widehat I_5 - I_1 I_2) }{dt} & \simeq \left( 2 a_0 +2 b_0
+ \frac{4 a_1}{3} \, I_1  - \frac{2 b_2}{3} \, I_1 \right) (\widehat I_5 - I_1 I_2) \,,
\end{align}
and 
\begin{align}
\frac{d\widehat I_6}{dt} & = 6 a_0 \, \widehat I_6 \,,
&&
\frac{d\widehat I_7}{dt} \simeq \left( 4 a_0 + 2 b_0 + \frac{2 a_1-2 b_2}{3} \, I_1 \right) \widehat I_7 
\,,
\cr 
\frac{d\widehat I_8}{dt} & = 6 b_0 \, \widehat I_8 \,,
&&
\frac{d\widehat I_9}{dt}  \simeq \left( 4 b_0 + 2 a_0 + \frac{4 a_1-4 b_2}{3} \, I_1 \right) \widehat I_9  \,,
\end{align}
and
\begin{align}
\frac{d (\widehat I_{10}- \widehat I_3 \widehat I_4)}{dt} &\simeq \left( 4 a_0 + 4 b_0 + \frac{2 a_1+2 b_2}{3} \, I_1 \right) (\widehat I_{10} - \widehat I_3 \widehat I_4)\,.
\end{align}
together with
\begin{align}
\frac{d \widehat I_{11}^-}{dt} &\simeq  \left( 6 a_0 + 6 b_0 + 2 a_1 \, I_1 \right) I_{11}^- \,.
\end{align}
As in the 2G example, only the coefficients $a_0,b_0,a_1,b_2$ in 
(\ref{3G-1loop}) are needed in this approximation.
Solving for the latter, one obtains
\begin{align}
  a_0 \, dt = -\frac{d I_1}{6 \,I_1} + \frac{d\widehat I_3}{3 \, \widehat I_3}   \,, \qquad 
 &  a_1 \, dt \simeq \frac{1}{2I_1}\left( 2 \, \frac{dI_1}{I_1} 
 -\frac{d\widehat I_3}{\widehat I_3} \right) \,, 
  \cr  
  b_0 \, dt = -\frac{d I_2}{6 \,I_2} + \frac{d\widehat I_4}{3 \, \widehat I_4} \,, 
  \qquad &
 b_2 \, dt  \simeq  \frac{1}{2 I_1} \left(
  2 \, \frac{dI_2}{I_2}
 -  \frac{d\widehat I_4}{\widehat I_4} 
    \right) \,. 
\end{align}
This leaves 7 relations that can be used to identify 
RG-invariant combinations of flavour invariants,
\begin{align}
\frac{d\widehat I_6}{\widehat I_6} +
\frac{dI_1}{I_1} = 2 \,  \frac{d\widehat I_3}{\widehat I_3}  \,,
\qquad 
\frac{d\widehat I_8}{\widehat I_8} + \frac{dI_2}{I_2} \simeq 
2 \, \frac{d\widehat I_4}{\widehat I_4} \,,
\end{align}
and
\begin{align}
\frac{d(\widehat I_5 - I_1 I_2)}{\widehat I_5 - I_1 I_2} +  \frac{dI_2}{I_2} 
&\simeq   \frac{dI_1}{I_1}  +  \frac{d\widehat I_4}{\widehat I_4}
\,,
\end{align}
and 
\begin{align}
 \frac{d\widehat I_7}{\widehat I_7} +  \frac{dI_2}{I_2}  &\simeq
 \frac{d \widehat I_3}{\widehat I_3}
+ \frac{d\widehat I_4}{\widehat I_4} \,,
\qquad  
\frac{d\widehat I_9}{\widehat I_9} + 2 \, \frac{dI_2}{I_2} \simeq
\frac{dI_1}{I_1} + 2 \, \frac{d\widehat I_4}{\widehat I_4} \,,
\end{align}
and
\begin{align}
 \frac{d(\widehat I_{10}- \widehat I_3 \widehat I_4)}{\widehat I_{10} - \widehat I_3 \widehat I_4 } 
 &\simeq 
\frac{d\widehat I_3}{\widehat I_3} + \frac{d\widehat I_4}{\widehat I_4} 
\,,
 \qquad
 \frac{d I_{11}^-}{I_{11}^-} +  \frac{dI_2}{I_2}  \simeq
  \frac{dI_1}{I_1} + \frac{d\widehat I_3}{\widehat I_3} 
 + 2 \, \frac{d\widehat I_4}{\widehat I_4} \,.
\end{align}
Here each of the invariants 
is to be read as a function of $(I_1,I_2,\widehat I_3,\widehat I_4)$. 
As in the 2G case, the relations can be easily integrated, resulting in
\begin{align}
& \frac{I_1 \widehat I_6}{(\widehat I_3)^2} \simeq \text{const.} \simeq \frac{y_u^2}{y_c^2} 
\sim \epsilon^5 \,,
\qquad 
\frac{I_2 \widehat I_8}{(\widehat I_4)^2} \simeq \text{const.} \simeq 
\frac{y_d^2}{y_s^2}  \sim \epsilon^2 \,,
\label{set1}
\end{align}
and 
\begin{align}
&  \frac{I_1 \widehat I_4}{I_2 (\widehat I_5 - I_1 I_2)} 
  \simeq \text{const.} \simeq \frac{y_s^2}{y_b^2 \, |V_{cb}|^2} \sim \epsilon^2 \,,
\end{align}
and 
\begin{align}
&  \frac{\widehat I_3 \widehat I_4}{I_2 \widehat I_7} 
  \simeq \text{const.} \simeq \frac{y_s^2}{y_b^2 \, |V_{ub}|^2} \sim \epsilon \,, 
  \qquad 
  \frac{I_1 \widehat I_4^2}{I_2 \widehat I_9} 
  \simeq \text{const.} \simeq \frac{y_s^2}{y_b^2 \, |V_{td}|^2} \sim \epsilon \,,
\end{align}
and 
\begin{align}
\frac{\widehat I_{10}- \widehat I_3 \widehat I_4}{\widehat I_3 \widehat I_4} 
  \simeq \text{const.} \simeq  |V_{us}|^2 \sim \epsilon \,,
  \qquad 
\frac{I_1\widehat I_3 \widehat I_4}{I_2 I_{11}^-} \simeq \text{const.} 
\simeq \frac{4 y_s^2}{3 y_b^2 \, {\rm Im}[ V_{ud} V_{ub}^* V_{td}^* V_{tb}]} \sim \epsilon \,.
\label{set2}
\end{align}
This explicitly shows, how the known simplifications for the RG solutions 
of quark masses and mixing angles that 
arise in the limit of large top-quark Yukawa coupling 
(see also \cite{Liu:2009vh}) can be translated to 
the set of flavour invariants in a straightforward manner.

\subsubsection{One-loop Solutions in the SM}

As in the 2G-case, we can derive explicit solutions to the 
RG equations, using the one-loop expressions for the coefficients in
Table~\ref{tab:aibi} and the approximations from the 
Yukawa hierarchies discussed in the previous paragraph.
With our definitions of flavour invariants, 
the approximate RG equations for the invariants $I_{1,2}$
and $\widehat I_{3,4}$ looks identical to the 2G case
in (\ref{I1SM},\ref{I2SM},\ref{I345SM}).
As a consequence, we can again express the running of 
the 11 invariants in terms of the RG function $G(\mu,\mu_0)$
defined in (\ref{Gfunc}) from the evolution of the leading
invariant 
\begin{align}
 I_1(\mu) &\simeq 
  \eta^{8/\beta_0}\, I_1(\mu_0) \, G(\mu,\mu_0) \,, \qquad \eta= \frac{\alpha_s(\mu)}{\alpha_s(\mu_0)} \,.
\end{align}
Defining normalized invariants as before (using a slightly different notation), 
we have
\begin{align}
x_2(\mu) \equiv \frac{I_2(\mu)}{I_1(\mu)}
  &\simeq\left[G(\mu,\mu_0)\right]^{-2/3} x_2(\mu_0) 
 \,,
 \cr 
x_3(\mu) \equiv \frac{\widehat I_3(\mu)}{(I_1(\mu))^2}
  &\simeq  \left[G(\mu,\mu_0)\right]^{-1/3} x_3(\mu_0)
 \,,
 \cr 
x_4(\mu) \equiv \frac{\widehat I_4(\mu)}{(I_2(\mu))^2}
  &\simeq  \left[G(\mu,\mu_0)\right]^{+1/3} x_4(\mu_0)
 \,. 
\end{align}
and the remaining scaling relations follow from (\ref{set1}-\ref{set2}).
In this way, we 
recover the results for the approximate RG running of 
CKM mixing angles as discussed in \cite{Balzereit:1998id}.

\paragraph{Comparison with Harrison et al.}

In a paper by Harrison et al.~\cite{Harrison:2010mt}
it has been highlighted that, within the SM, the one-loop 
RG equations exhibit two combinations of flavour invariants
that are stable with respect to RG flow,
\begin{align}
 \frac{d}{dt} \left( \frac{\tr[UD]}{\left(\det[UD]\right)^{1/3}} \right) 
 = \frac{d}{dt} \, \frac{\widehat I_5}{(\widehat I_6 \widehat I_8)^{1/3}} &= 0 \qquad \mbox{(SM@1-loop)} \,, \\
 \frac{d}{dt} \left( \tr\left[ (UD)^{-1} \right] \left(\det[UD]\right)^{1/3} \right)
    = \frac{d}{dt} \, 
    \frac{\widehat I_{10}}{(\widehat I_6 \widehat I_8)^{2/3}} & = 0 \qquad \mbox{(SM@1-loop)} \,.
\end{align}
In our notation, we have 
\begin{align}
  \frac{d}{dt} \, \frac{\widehat I_5}{(\widehat I_6 \widehat I_8)^{1/3}}
 & = \frac{ 
   2 \, (a_1+b_2) \, \widetilde I_7 
 + 2 \, (a_2+b_1) \, \widetilde I_9}{ \left(\det[UD]\right)^{1/3}} \,,
\\[0.2em]
  \frac{d}{dt}
 \, 
    \frac{\widehat I_{10}}{(\widehat I_6 \widehat I_8)^{2/3}} 
 & = \frac{2 \, (a_1+b_2) \left( I_1 \widehat I_{10} - 3 \widehat I_4 \widehat I_6\right)
+ 2 (a_2+b_1) \left( I_2 \widehat I_{10} - 3 \widehat I_3 \widehat I_8\right) }{3 \left( \det[UD] \right)^{2/3}} 
  \,. 
\end{align}
This indeed vanishes for $a_1=-b_2=-a_2=b_1$ which holds within the SM, see Table~\ref{tab:aibi}.


\section{Summary and Outlook}
\label{summary}

From the experimental 
as well as form the theoretical side (see e.g.\ the reviews in 
\cite{Antonelli:2009ws,Bediaga:2012py,Buras:2013ooa,Buras:2010wr}), the quark flavour physics 
program is currently entering the precision era. 
The goal is to find hints to physics beyond the Standard Model (SM) 
from dedicated experiments, notably LHCb and BELLE~II. 
Still, the answer to the flavour puzzle itself may reside 
at extremely high scales, possibly as high as the Planck scale. 
In any case, the determination of flavour observables occurs at low energies, 
and thus for any comparison with ``new physics'' models one needs to 
include the renormalization-group (RG) running of the flavour parameters in 
a given theoretical framework. 
In principle, there are various roads to discuss this. 
On the one hand,  one can consider the entries of the $3\times 3$ Yukawa matrices 
and study their RG evolution; but these depend on an arbitrary choice 
of basis in flavour space. 
Alternatively, one can use the physical parameters, i.e.\ the six quark masses 
together with four independent CKM parameters to describe quark mixing and CP violation
in weak interactions; but these have rather complicated relations to the Yukawa couplings.

In this paper, we have chosen an intermediate point of view 
and considered simple combinations of Yukawa couplings that are 
independent of the orientation of the flavour basis. 
In terms of these flavour invariants we have formulated RG equations 
which are basis independent and allow for a transparent implementation 
of flavour hierarchies as observed in the SM or its minimal-flavour-violating
(MFV) extensions.
Expanding systematically in small parameters, 
we have also constructed simple analytic solutions 
for the RG evolution of a set of polynomially independent
flavour invariants.

Discussing the RG flow in terms of flavour invariants may be advantageous 
to discuss models with dynamical flavour symmetry breaking, where
the Yukawa couplings emerge as vacuum expectation values (VEVs)
of some scalar flavon fields. The scalar potential generating these VEVs 
will be constructed in terms of polynomials of 
flavour invariants of a given canonical mass dimension. 
In MFV-like constructions (see e.g.\ \cite{Albrecht:2010xh}), 
these can be reduced to the set of invariants discussed in this work.
More complicated situations arise if one implements the spontaneous breaking 
of a gauged flavour symmetry on the level of renormalizable interactions. 
This leads to an ``inverted-MFV'' scenario, 
where the fundamental flavour invariants are approximately
given as polynomials of the inverse Yukawa matrices \cite{Grinstein:2010ve}. 
Even more complicated relations can arise in a recently proposed model 
with dynamical flavour-symmetry breaking with a unification scheme 
according to Pati and Salam \cite{Feldmann:2015zwa}. While the 
general form of the RG equations (\ref{eq:3Grge-generic}) will remain
the same, the coefficients will have a more complicated dependence 
than in MFV scenarios.
In any of these cases, the renormalization-group flow of the invariants is 
needed to constrain the theoretical NP parameters at a high scale 
from  flavour observables at low scales, and eventually give us some clue 
on the solution of the flavour problem.

\section*{Acknowledgements}

This work is supported 
by the Deutsche Forschungsgemeinschaft (DFG) within Research Unit FOR 1873
(``Quark Flavour and Effective Field Theories'').

\vskip1em
\hrulefill 
\vskip1em

\appendix

\section{Cayley-Hamilton Identities}

\label{app:CH}

\subsection{Two-Generation Case }

The Cayley-Hamilton identity for $2\times 2$ matrices $M$ reads
\begin{align}
 0 &= M^2 - \tr[M] \, M +  \det M \, \mathds{1} \,.
\label{cl2}
\end{align}
Taking the trace and solving for $\det M$, one obtains
\begin{align}
 \det M &= \frac12 \left( \tr^2[M]-\tr[M^2] \right) \,.
\end{align}
Multiplying (\ref{cl2}) with $M^{-1}$, and solving for $\adj M = M^{-1} \det M$, one obtains
\begin{align}
\adj M& = \tr[M]\, \mathds1 - M \qquad \Rightarrow \qquad 
\tr[\adj M] =  \tr[M] \,.
\label{adjM2}
\end{align}
Inserted back into (\ref{cl2}) yields
\begin{align}
 M^2
 &= \tr[M] \, M  - \det [M] \, \mathds{1} \,.
\end{align}
For traceless matrices, this further simplifies to
\begin{align}
 M^2 &= \frac12 \, \tr[M^2]  \, \mathds{1}\qquad (\tr[M]=0)\,.
\end{align}
Therefore any power of $2\times 2$ matrices $M$ can be reduced to the basis 
$\{\mathds{1},M \}$ with coefficients built from  polynomials of $\tr M$, $\tr M^2$
which are invariant under \emph{unitary} basis transformations.

For matrices $Y$ which transform under \emph{bi-unitary} transformations,
Eq.~(\ref{adjM2}) generalizes to
\begin{align}
 \adj Y &= \frac{\det[Y]}{\det[M]} \,Y^\dagger
 \left(\tr[M]\, \mathds{1} -  M \right) 
 \qquad \mbox{($M\equiv Y Y^\dagger$)} \,.
\end{align}

\subsection{Three-Generation Case}

The Cayley-Hamilton identity for $3\times 3$ matrices $M$ reads
\begin{align}
 0 &= M^3 - \tr[M] \, M^2 + \frac12 \left( \tr^2[M]- \tr[M^2] \right) M - \det M \, \mathds{1} \,.
\label{cl}
\end{align}
Taking the trace and solving for $\det M$, one obtains
\begin{align}
 \det M &= \frac13 \left( \tr[M^3]-\frac32 \, \tr[M]\tr[M^2]+\frac12 \, \tr^3[M] \right) \,.
\end{align}
Multiplying (\ref{cl}) with $M^{-1}$, and solving for $\adj M = M^{-1} \det M$, one obtains
\begin{align}
\adj M& = M^2 - \tr[M] \, M + \frac{1}{2} \left(\tr^2[M] - \tr[M^2]\right) \mathds1 \cr 
\Rightarrow \qquad \tr[\adj M] &=  \frac{1}{2} \left(\tr^2[M] - \tr[M^2]\right) \,.
\label{adjM3}
\end{align}
Inserted back into (\ref{cl}) yields
\begin{align}
 M^3 
 &= \tr[M] \, M^2 - \tr [\adj M] \, M + \det [M] \, \mathds{1} \,.
\end{align}
For traceless matrices, this further simplifies to
\begin{align}
 M^3 &= \frac12 \, \tr[M^2] \, M  + \frac13 \, \tr[M^3]  \, \mathds{1}\qquad (\tr[M]=0)\,.
\end{align}
Therefore any power of $3\times 3$ matrices $M$ can be reduced to the basis 
$\{\mathds{1},M,M^2 \}$ with coefficients built from invariants 
that are polynomials of $\tr M$, $\tr M^2$, $\tr M^3$.

Similarly as before, 
for matrices $Y$ which transform under \emph{bi-unitary} transformations,
Eq.~(\ref{adjM3}) generalizes to
\begin{align}
 \adj Y &= \frac{\det[Y]}{\det[M]} \, Y^\dagger
 \left(\tr[\adj M]\, \mathds{1} -  \tr[M] \, M + M^2 \right) 
 \qquad \mbox{($M\equiv Y Y^\dagger$)} \,.
\end{align}


\section{3G Flavour Invariants, Yukawa Couplings and CKM Elements}

\label{translate}

For our convention to define 10+1 polynomially independent flavour invariants, 
the explicit expressions in terms of Yukawa couplings and mixing
angles read as follows.
The quadratic invariants are
\begin{align}
  I_1 &\equiv \tr[U] = \sum_{i=u,c,t} y_i^2 \,, \qquad
  I_2 \equiv \tr[D] =  \sum_{j=d,s,b} y_j^2 \,.
\end{align}
Again, $I_1$ and $I_2$ quantify the overall size of flavour-symmetry
breaking in the up- and down-quark sector, respectively.
Quartic invariants appear as
\begin{align}
  \widehat I_3 &\equiv \tr[\adj U] = \sum_{i=u,c,t} \widetilde y_i^2 \,, \qquad 
  \widehat I_4 \equiv \tr[\adj D] = \sum_{j=d,s,b} \widetilde y_j^2 \,, 
\end{align}
and 
\begin{align}
  \widehat I_5 &\equiv \tr[U D] = 
  \sum_{i=u,c,t} \sum_{j=d,s,b} y_i^2 y_j^2 \, |V_{ij}|^2  \,,
\end{align}  
where we have defined $\widetilde y_u^2 = y_c^2 y_t^2$, $\widetilde y_d^2 = y_s^2 y_b^2$ etc.
Continuing with the 
sixth-order invariants, we have
\begin{align}
  \widehat I_6& = \tr[U \, \adj U] = 3 \det U = 3 \, y_u^2 y_c^2 y_t^2 \,, 
  \qquad 
  \widehat I_8 = \tr[D \, \adj D] = 3 \det D = 3 \, y_d^2 y_s^2 y_b^2 \,,
\end{align}
and
\begin{align}
  \widehat I_7 &\equiv \tr[\adj U \, D] = 
  \sum_{i=u,c,t} \sum_{j=d,s,b} \widetilde y_i^2 y_j^2 \, |V_{ij}|^2 \,, 
  \cr 
  \widehat I_9 &\equiv  \tr[U \, \adj D]= 
  \sum_{i=u,c,t} \sum_{j=d,s,b} 
   y_i^2 \widetilde y_j^2 \, |V_{ij}|^2 
\,.
\end{align}
The eight-order invariant reads
\begin{align}
 \widehat I_{10} \equiv \tr[ \adj U \, \adj D] =
  \sum_{i=u,c,t} \sum_{j=d,s,b} 
   \widetilde y_i^2 \widetilde y_j^2 \, |V_{ij}|^2 
 \,.
 \end{align}
Finally, the CP-odd invariant 
\begin{align}
I_{11}^- & \equiv \tr[A_8^3] = \frac34 \, 
(y_t^2-y_c^2)(y_t^2-y_u^2)(y_c^2-y_u^2)
(y_b^2-y_s^2)(y_b^2-y_d^2)(y_s^2-y_d^2) 
\,
{\rm Im}\left[ V_{ud} V_{ub}^* V_{td}^* V_{tb} \right]
\,,
\end{align}
is proportional to the Jarlskog determinant
\cite{Jarlskog:1985ht}.
 

\section{Two-Loop RG equations for 3G Flavour Invariants} 
 
 \label{app:3G2loop}

The two-loop approximation for RG equations of the quark 
Yukawa matrices $Y_U$ and $Y_D$ in 
(\ref{eq:3Grge-generic}) is obtained 
by keeping factors that are at most quartic 
in the Yukawa couplings, i.e.\ neglecting the contributions with
the flavour matrices $A_{U,D}$,
\begin{align}
 \frac{d Y_U}{d\ln\mu} &= 
\left( a_0 \, \mathds{1} 
+ a_1 \, U_8 + a_2 \, D_8 
+ a_3 \adj U_8 + a_4 \, \adj D_8 
+ a_5 \, S_8 + i a_6 \, A_8 + \ldots \right) Y_U(\mu) \,,
\cr
 \frac{d Y_D}{d\ln\mu} &= 
\left( b_0 \, \mathds{1} 
+ b_1 \, D_8 + b_2 \, U_8 
+ b_3 \, \adj D_8 + b_4 \, \adj U_8 
+ b_5 \, S_8 -i b_6 \, A_8 + \ldots \right) Y_D(\mu)
\,.
\label{3G-2loop}
\end{align} 
From this ansatz, it is straightforward -- though tedious -- to calculate 
the two-loop RG equations for the eleven flavour invariants.
First, we have
\begin{align}
  \frac{dI_{1}}{dt} & \simeq 2 a_{0} \, I_{1} + 2a_{1} \, I_{3} + 2a_{2} \, I_{5} 
   + 2 a_3 \, I_6 + 2 a_4 \, I_9 + 2 a_5 \, \widetilde I_7 \,,
  \cr 
  \frac{dI_{2}}{dt} & \simeq  2 b_{0} \, I_{2} + 2b_{1} \, I_{4} + 2b_{2} \, I_{5} 
   + 2 b_3 \, I_8 + 2 b_4 \, I_7 + 2 b_5 \, \widetilde I_9 \,,
\end{align}
where here and in the following the abbreviations for the combinations of
flavour invariants are the same as in Table~\ref{metric3G}.
Then 
\begin{align}
  \frac{d\widehat I_{3}}{dt} & \simeq 
  4 a_{0} \, \widehat I_{3} 
  - 2 a_{1} \, I_{6} 
  - 2 a_{2} \, I_{7} 
  - 2 a_{3} \, P_1 
  - 2 a_4 \, P_2 
  - 2 a_5 \, P_3 
  \,,
  \cr 
  \frac{d\widehat I_{4}}{dt} & \simeq
  4 b_{0} \, \widehat I_{4} 
  - 2 b_{1} \, I_{8} 
  - 2 b_{2} \, I_{9} 
  - 2 b_{3} \, P_4 
  - 2 b_4 \, P_2
  - 2 b_5 \, P_5 
\end{align}
and 
\begin{align}
  \frac{d\widehat I_{6}}{dt} & = 
  6 a_{0} \, \widehat I_{6}
  \,, \qquad 
  \frac{d\widehat I_{8}}{dt}  = 
  6 b_{0} \, \widehat I_{8} \,. 
\end{align}
For the mixed invariants, one obtains
\begin{align}
  \frac{d\widehat I_5}{dt} & \simeq
  (2 a_0 + 2 b_0) \, \widehat I_5
  + (2 a_1 + 2 b_2) \, \widetilde I_7 
  + (2 a_2 + 2 b_1) \, \widetilde I_9
  \cr & \quad 
  + (2 a_3 + 2 b_4) \, P_3 
  + (2 a_4 + 2 b_3) \, P_5
  + (2 a_5 + 2 b_5) \, I_{10}
  + (2 a_6 + 2 b_6) \, \widetilde I_{10} \,,
\end{align}
and
\begin{align}
  \frac{d\widehat I_7}{dt} & \simeq
  (4 a_0 +2 b_0) \, \widehat I_7 
  + (2 a_1- 2 b_2) \left( \frac{I_1 \widehat I_7}{3}- I_2 \widehat I_6  \right)
 - (2 a_2 - 2 b_1) \left(
 \widehat I_{10} - \widehat I_3 \widehat I_4 + \frac{2 I_2 \widehat I_7}{3}
 \right) 
  \cr 
  & \quad 
  + (2 a_3- 2 b_4) \left( 2 \widehat I_5 \widehat I_6 - 3 I_5 \widehat I_6 - \frac{2\widehat I_3 \widehat I_7}{3} \right)
  + (2 a_4- 2 b_3) \left( \frac{\widehat I_4 \widehat I_7}{3} - \widehat I_3 \widehat I_8 \right)
  \cr 
  & \quad 
  + (2 a_5-2 b_5) \left( (2\widehat I_4 - I_2^2) \, \widehat I_6 
  + \frac{\widehat I_5 \widehat I_7}{3} + Q_1 \right)
  + (2a_6+ 2b_6) \, Q_1
  \,, 
  \cr 
  \frac{d\widehat I_9}{dt} & \simeq
  (2 a_0 +4 b_0) \, \widehat I_9
  - (2 a_2- 2 b_1) \left( \frac{I_2 \widehat I_9}{3}- I_1 \widehat I_8  \right)
  - (2 b_2- 2 a_1) \left(
 \widehat I_{10} - \widehat I_3 \widehat I_4 + \frac{2 I_1 \widehat I_9}{3} \right)
  \cr 
  & \quad 
  - (2 a_4- 2 b_3) \left( 2 \widehat I_5 \widehat I_8 - 3 I_5 \widehat I_8 - \frac{2\widehat I_4 \widehat I_9}{3} \right)
  - (2 a_3- 2 b_4) \left( \frac{\widehat I_3 \widehat I_9}{3} - \widehat I_4 \widehat I_6 \right)
  \cr 
  & \quad 
  - (2a_5-2b_5) \left( (2\widehat I_3 - I_1^2) \, \widehat I_8 
  + \frac{\widehat I_5 \widehat I_9}{3}- Q_2 \right)
  - (2 a_6+ 2 b_6) \, Q_2
  \,, 
\end{align}
and 
\begin{align}
  \frac{d\widehat I_{10}}{dt} & \simeq
  (4 a_0 + 4  b_0) \, \widehat I_{10}
  + (2 a_1 + 2 b_2) \left( \frac{I_1 \widehat I_{10}}{3} -  \widehat I_4 \widehat I_6 \right)
  + (2 a_2 + 2 b_1) \left( \frac{I_2 \widehat I_{10}}{3} -  \widehat I_3 \widehat I_8 \right)
  \cr 
  & \quad 
  - (2 a_3 + 2 b_4) \left( \widehat I_6 \, (\widehat I_9 - I_1 \widehat I_4) 
   + \frac{2 \widehat I_3 \widehat I_{10}}{3} \right)
   - (2 a_4 + 2 b_3) \left( (\widehat I_7 - I_2 \widehat I_3 ) \, \widehat I_8  
   +\frac{2 \widehat I_4 \widehat I_{10}}{3} \right)
   \cr 
   & \quad 
   + (2a_5 + 2b_5)
   \left( \frac{\widehat I_5 \widehat I_{10}}{3} - 3 \widehat I_6 \widehat I_8 - Q_4\right)
    - (2 a_6 +  2 b_6) \, Q_4 \,.
\end{align}
and
 \begin{align}
  \frac{dI_{11}^-}{dt} & \simeq 
  \left(6 a_0 + 6 b_0 + 
    2 a_1  I_1 + 2 b_2  I_2  - 2 a_4 \widehat I_3 - 2 b_4 \widehat I_4 \right) I_{11}^-
    \cr & \quad 
    + \left( 
    (a_5+b_5) \, \frac{\widehat I_5 + I_1 I_2}{2}
    - (a_6+b_6) \, I_5 \right) I_{11}^- \,. 
 \end{align}

\vskip1em
\hrulefill 

\end{document}